\documentclass[a4paper,11pt]{article}
\pdfoutput=1
\usepackage{jcappub}
\usepackage[below]{placeins}

\newcommand{\dd}{\mathrm{d}}
\newcommand{\dm}[1]{d_{m=#1}}
\newcommand{\dLz}{d_L(z)}
\newcommand{\Nws}{N_{\text{ws}}}
\newcommand{\ns}{n_s}
\newcommand{\nzero}{n_0}
\newcommand{\Lnu}{L_{\nu}}
\DeclareMathOperator*{\argmax}{arg\,max}

\title{Detection prospects for high energy neutrino sources from the anisotropic matter distribution in the local Universe}

\author[a]{Philipp Mertsch,}
\author[a]{Mohamed Rameez}
\author[a]{and Irene Tamborra}
\affiliation[a]{Niels Bohr International Academy, Niels Bohr Institute, Blegdamsvej 17, 2100 Copenhagen, Denmark}
\emailAdd{mertsch@nbi.ku.dk}
\emailAdd{mohamed.rameez@nbi.ku.dk}
\emailAdd{tamborra@nbi.ku.dk}

\abstract{Constraints on the number and luminosity of the sources of the cosmic neutrinos detected by IceCube have been set by targeted searches for point sources. We set complementary constraints by using the 2MASS Redshift Survey (2MRS) catalogue, which maps the matter distribution of the local Universe. Assuming that the distribution of the neutrino sources follows that of matter, we look for correlations between ``warm'' spots on the IceCube skymap and the 2MRS matter distribution. Through Monte Carlo simulations of the expected number of neutrino multiplets and careful modelling of the detector performance (including that of IceCube-Gen2),  we demonstrate that sources with local density exceeding $10^{-6} \, \text{Mpc}^{-3}$ and neutrino luminosity $L_{\nu} \lesssim 10^{42} \, \text{erg} \, \text{s}^{-1}$ ($10^{41} \, \text{erg} \, \text{s}^{-1}$) will be efficiently revealed by our method using IceCube (IceCube-Gen2). At low luminosities such as will be probed by IceCube-Gen2, the sensitivity of this analysis is superior to requiring statistically significant direct observation of a point source.
}

\begin{document}
\maketitle
\flushbottom

\section{Introduction}
\label{sec:introduction}

The detection of cosmic neutrinos with energies up to few PeV has marked the beginning of high-energy neutrino astronomy~\cite{Aartsen:2013bka,Aartsen:2013jdh,Aartsen:2014gkd,Aartsen:2014muf,Aartsen:2015knd,Aartsen:2015rwa}. The observed high-energy neutrino events are consistent with an isotropic distribution in the sky. Their flavor content is compatible with an equal distribution among $\nu_{\text{e}}$, $\nu_{\mu}$ and $\nu_{\tau}$, and their intensity is consistent with the Waxman-Bahcall bound~\cite{Waxman:1998yy}. These features hint towards a flux of astrophysical origin~\cite{Anchordoqui:2013dnh,Halzen:2013dva,Ahlers:2015lln}. 

Several sources have been discussed as possible emitters of the high-energy neutrinos detected by IceCube, although none of them seems to explain the observed flux with high significance. Indeed the principal challenge in identifying the sources of the diffuse flux lies in the fact that most of these events, being cascade events, have angular resolutions of up to $\sim 15^{\circ}$~\cite{Aartsen:2014gkd,Aartsen:2015zva}.

Galactic sources have been proposed in the literature as high-energy neutrino factories and are currently not excluded as an explanation for the IceCube neutrino events~\cite{Tjus:2015rck}, although they cannot explain the whole observed flux. The largest contribution should come from extragalactic sources,  the ones mostly discussed in the literature are cosmic-ray reservoirs such as star-forming galaxies and galaxy clusters or groups~\cite{Waxman:2015ues,Murase:2013rfa,Tamborra:2014xia,Ando:2015bva,Senno:2015tra,Chakraborty:2015sta,Chang:2014sua,Fang:2016amf,Zandanel:2014pva}; gamma-ray bursts~\cite{Meszaros:2015krr}, active galactic nuclei and blazars~\cite{Murase:2015ndr,Neronov:2016ksj,Padovani:2015mba} as well as low-power or hidden cosmic accelerators~\cite{Murase:2013ffa,Murase:2015xka,Senno:2015tsn,Tamborra:2015qza,Tamborra:2015fzv,Wang:2015mmh,Dai:2016gtz,Senno:2016bso}.  

It is very likely that the observed energy spectrum is a superposition of the diffuse emission coming from several sources. This could be corroborated if spectral breaks were to become visible with increasing statistics~\cite{Kowalski2016}. Note also that even assuming that the observed neutrino flux is mostly described by a single power law, existing analysis are in mild tension on the exact value of the spectral index according to the event sample adopted in the analysis~\cite{Aartsen:2015knd,Aartsen:2016xlq}. 

At the same time, the IceCube collaboration has carried out targeted searches for the detection of point sources (PS), see e.g.~Refs.~\cite{Aartsen:2016oji,Aartsen:2016tpb,Aartsen:2016qcr,Adrian-Martinez:2015ver,Aartsen:2015yva},  up to now with negative results, which is not in contradiction with theoretical expectations~\cite{Ahlers:2014ioa,Murase:2016gly,Feyereisen:2016fzb,Halzen:2016seh}. In principle, dedicated searches for astrophysical PS of neutrinos are  more sensitive when performed with samples of track like events~\cite{Aartsen:2016oji} from charged current (CC) interactions of $\nu_{\mu}$ and $\bar{\nu}_{\mu}$ of astrophysical origin. Those samples are however dominated by muons from CC interactions of atmospheric muon neutrinos in the Northern sky, and muons produced in interactions of cosmic rays with the upper atmosphere in the Southern sky. Although background dominated, such samples have significantly larger effective volumes, since the outer layers of the detector are not used as a veto and events with interaction vertices far away from the instrumented volume can still be detected due to the large range of the muon in ice at high energies. The arrival direction of the neutrino can be estimated with less than $1^{\circ}$ precision~\cite{Aartsen:2016oji}, and astrophysical PS of neutrinos can be identified by looking for statistically significant clusters of high-energy events, over the atmospheric backgrounds.

The most general IceCube search targeting steady PS{\footnote{We define a  ``steady'' PS as having no significant variability during the IceCube data taking period, i.e.~7 years in our case.} of neutrinos anywhere in the sky is the all sky PS scan~\cite{Aartsen:2016oji}. This search is sensitive to PS with $E_\nu^{-2}$ fluxes as low as $E_\nu^2 \Phi_{\nu_{\mu} + \bar{\nu}_{\mu}}\sim 4 \times 10^{-13} \, \text{TeV} \, \text{cm}^{-2} \, \text{s}^{-1}$. However, only PS producing fluxes higher than $E_\nu^2 \Phi_{\nu_{\mu} + \bar{\nu}_{\mu}}\sim 2\times 10^{-12} \, \text{TeV} \, \text{cm}^{-2} \, \text{s}^{-1}$ can be distinguished from a coincidental clustering of atmospheric neutrinos due to the large statistical trial factor involved in scanning every point in the sky. The distances (in equivalent redshift $z$) at which sources of a given muon neutrino luminosity $L_\nu$ can generate the above flux values are shown in the left panel of Fig.~\ref{fig:ZvsLum}. Sources with fluxes in between these values may produce clusters of events with high pre-trial significances, known as ``warm'' spots.

Note that targeted catalog and stacking searches using information about source directions from other astronomical messengers can constrain the $\nu_{\mu} + \bar{\nu}_{\mu}$ flux from one or more of these specific directions in the sky to levels even below the sensitivity of the all sky scan~\cite{Aartsen:2016lir, Aartsen:2014cva}. Note, however, as these constraints are valid only along  specific angular directions and do not constrain PS anywhere else in the sky.

The number of neutrino warm spots at different threshold pre-trial significances has been found to be consistent with background expectations, although a small excess exists in both the Northern and Southern skies corresponding to pre-trial significances with p-value $\sim 3\times10^{-3}$~\cite{Aartsen:2016oji}. If some of these warm spots are generated by neutrinos from nearby astrophysical sources, then  the  directions of these warm spots must correlate with the anisotropic matter distribution in the local Universe.

\begin{figure}[tbh]
\centering
\includegraphics[width=0.49\textwidth]{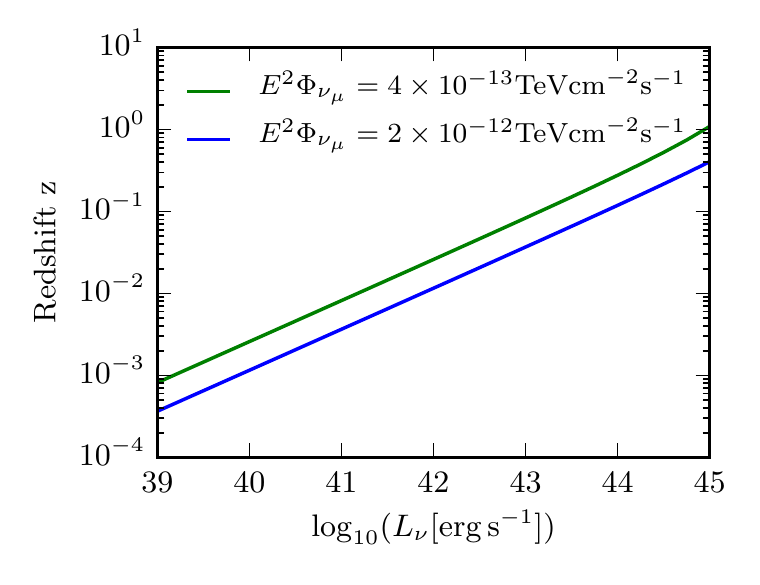}
\includegraphics[width=0.49\textwidth]{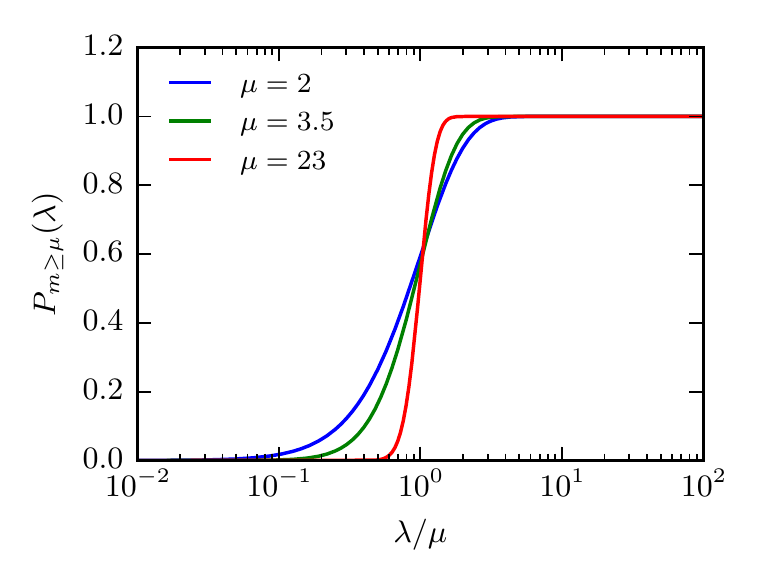}
\caption{\textbf{Left:} Redshift $z$ at which a source of given luminosity $\Lnu$ and with an energy spectrum $\propto E_\nu^{-2}$  will produce a $\nu_{\mu}$ flux corresponding to the benchmark sensitivities of the IceCube PS all sky scan as from Ref.~\cite{Aartsen:2016oji}. Fluxes in between the blue and the green lines will produce warm spots. \textbf{Right:} Probability that a source at redshift $z$ produces a number of events $m \ge \mu$, see Eq.~\ref{eq:Pm}. Sources with $\lambda = (\dm{1} / \dLz)^2 \le \mu$ contribute negligibly to $N_{m \geq \mu}^{z \leq z_c}$, see Eq.~\ref{eqn:multiplet1}.}
\label{fig:ZvsLum}
\end{figure}

The above arguments motivate looking for correlations between the directions of the IceCube warm spots defined by the PS searches and the matter distribution in the nearby Universe. However, currently the positions and properties of the warm spots from the IceCube PS search are not publicly available. We therefore determine the sensitivity of this proposed correlation study, both in an analytical approach and through a detailed Monte Carlo study. To this end, we adopt the 2MASS Redshift Survey (2MRS) catalogue~\cite{Crook:2006sw} which describes the local distribution of matter out to z = 0.03. By looking for correlations between the directions of warm spots  (simulated with similar properties as observed by IceCube in its all sky PS scan~\cite{Aartsen:2016oji}) and the anisotropies of the local Universe as from the 2MRS catalogue, we derive constraints on the local density of neutrino sources as a function of their neutrino luminosity. We show that at low enough luminosities ($L_\nu < 10^{42}$ erg s$^{-1}$), these constraints can be more stringent than those derived from the non observation of any statistically significant PS, if anisotropies observed in 2MRS serve as a tracer of the distribution of the cosmic accelerators emitting neutrinos. We also discuss prospects for detecting point sources in IceCube-Gen2~\cite{Aartsen:2014njl}.
    
The paper is organised as follows. In order to support our analysis, in  Sec.~\ref{sec:analytical} we derive analytical estimates on the non--detection of a neutrino PS by IceCube as well as sensitivity constraints from correlating the distribution of warm spots with a tracer of the matter distribution in the nearby Universe. In Sec.~\ref{sec:2MRS}, we introduce the  2MRS catalogue that we will use as tracer of the local matter distribution and characterise its local composition. Section~\ref{sec:method} describes our analysis method as well as our simulated neutrino event sample. The main results of our work are presented in Sec.~\ref{sec:results} together with a forecast for IceCube-Gen2 and a discussion on future prospects of detecting neutrino PS. Outlook and conclusions are presented in Sec.~\ref{sec:conclusions}. Supplementary material to further clarify our analysis is reported in Appendixes~\ref{sec:averageTS}, \ref{sec:limits} and \ref{sec:n_vs_l}.

\section{Bounds on the local source density}
\label{sec:analytical}

In the following, we analytically compute estimates for two kinds of constraints on the local density of sources as a function of luminosity: a limit derived from the non--observation of significant hot spots by IceCube~\cite{Aartsen:2016oji} (``PS limit'') and the sensitivity from searching for a correlation between the distribution of warm spots and the distribution of matter in the nearby Universe (``correlation sensitivity'').

\subsection{Number of multiplet events from neutrino sources}
\label{sec:analytical1}

\begin{figure}[tbh]
\centering
\includegraphics[scale=1]{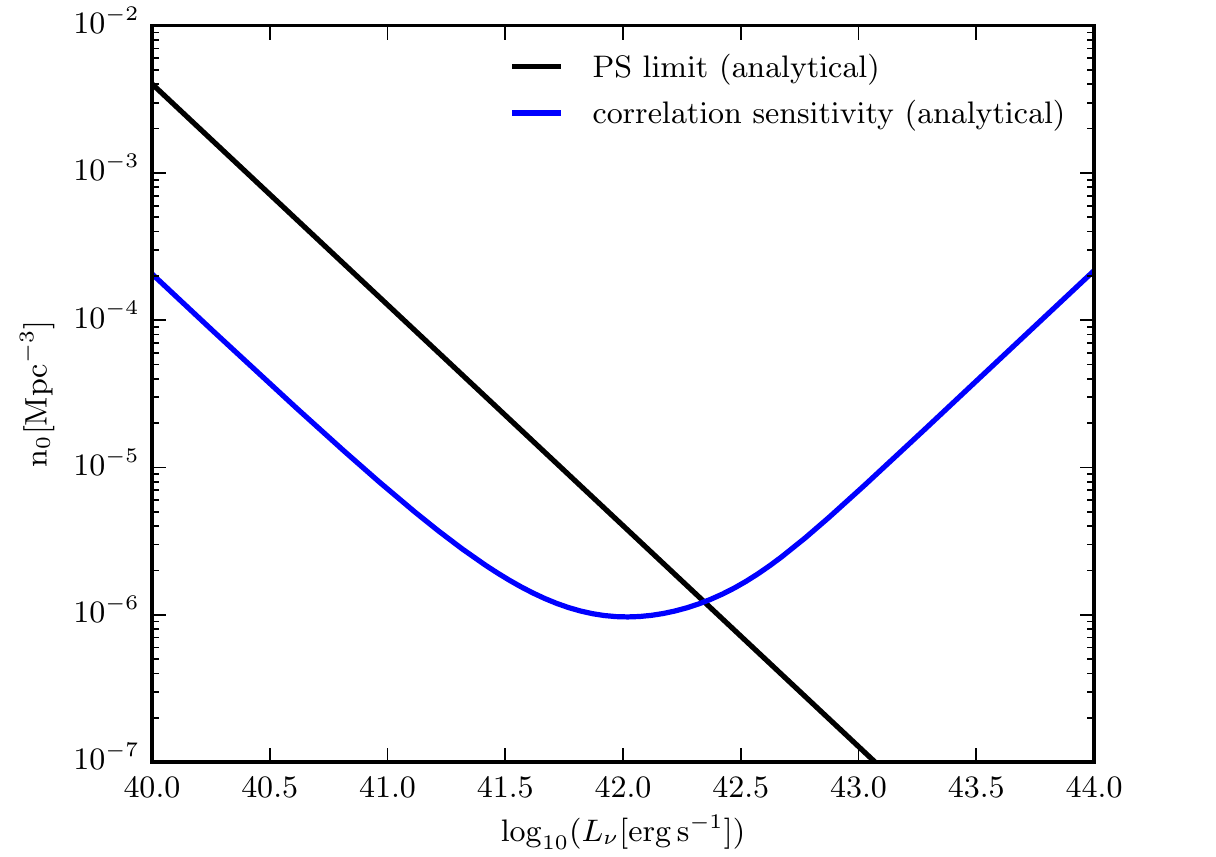}
\caption{Bounds on the local density $\nzero$ of neutrino sources as a function of luminosity $\Lnu$. The black line shows the bound from the non--observation of point sources (cf.\ Eq.~\ref{eqn:PSlimit}) for $\mu=23$. The blue line shows the sensitivity of the correlation analysis with the matter distribution in the local Universe (cf.\ Eq.~\ref{eqn:CorrSens}) for parameters $N_{\text{bkg}} = 30$, $\mu = 3$, $z_c = 0.02$, $\Delta \Omega  =4 \pi$ and $\sigma^2 = 2 \times 10^{-2}$. These parameters are not necessarily representative of the actual observational situation but merely for illustration purposes. For both bounds, we have assumed $\dm{1}$ as in Eq.~\ref{eqn:dm1_180Mpc}. For low luminosities, the correlation sensitivity is significantly smaller than the PS limit.}
\label{fig:limits}
\end{figure}

The number of neutrino sources at redshift $z$ can be written as the product of comoving source density $n_s(z)$ and differential comoving volume, $\dd \mathcal{V}(z)$. To compute the number of individual sources that IceCube would be able to detect or the number of warm spots, we also need the probability $P_{m \geq \mu} (\lambda)$ that a source at redshift $z$ produces a number of events $m$ equal or larger than $\mu$, $\lambda$ being the average number of events produced by the same source. Assuming that the neutrino events are uncorrelated, this is given by the complement of the cumulative Poisson distribution,
\begin{equation}
\label{eq:Pm}
P_{m \geq \mu}(\lambda) = \sum_{m=\mu}^{\infty} \frac{\lambda^m \mathrm{e}^{-\lambda}}{m !} = 1 - \frac{\Gamma(\mu, \lambda)}{\Gamma(\mu)} \, ,
\end{equation}
where $\Gamma(\mu)$ and $\Gamma(\mu, \lambda)$ are the gamma function and the incomplete gamma functions, respectively. In order to constrain the local source density for a given source luminosity as in Refs.~\cite{Murase:2012df,Murase:2016gly}, we have parametrised the expected number of events $\lambda$ from a source by its luminosity distance $\dLz$ and the luminosity distance $\dm{1}$ at which a source of a given luminosity $\Lnu$ would produce one neutrino event,
\begin{equation}
\lambda \equiv \left( \frac{\dm{1}}{\dLz} \right)^2 \, .
\label{eqn:lambda}
\end{equation}

We can thus write for the number of sources with redshift $z \leq z_c$ that contribute $\mu$ events or more~\cite{Murase:2016gly}
\begin{equation}
N_{m \geq \mu}^{z \leq z_c} = \nzero \Delta \Omega \int_0^{z_c} \mathrm{d} z \, \frac{(c/H_0) d_L^2(z)}{(1+z)^2 \sqrt{\Omega_m (1+z)^3 + \Omega_{\Lambda}}} \left(\frac{\ns(z)}{\nzero}\right) P_{m \geq \mu} (\lambda) \, , \label{eqn:multiplet1}
\end{equation}
with
\begin{equation}
\ns(z) = \nzero (1+z)^q \, ,
\label{eqn:SrcEvol}
\end{equation}
describing the source redshift evolution. The latter has been parametrised in the usual way: $q=3$ for star--formation--like evolution and $q=0$ for no evolution. For simplicity, we also assume that all sources belonging to the same class have the same luminosity (``benchmark sources''). Figure~\ref{fig:ZvsLum} (right panel)  shows the probability $P_{m \geq \mu} (\lambda)$ introduced in Eq.~\ref{eq:Pm} as a function of $\lambda/\mu$. We note that $P_{m \geq \mu} (\lambda)$ suppresses contributions to the redshift integral in Eq.~\ref{eqn:multiplet1} from sources with $\lambda \lesssim \mu$. 

Expanding $\dLz$ in $z$ up to $\mathcal{O}(z)$, $\lambda$ can be expressed as  
\begin{equation}
\lambda \simeq \left(\frac{\dm{1}}{(c/H_0) z}\right)^2 \simeq \frac{L_\nu}{10^{42} \, \text{erg} \, \text{s}^{-1}} \left(\frac{z}{0.042}\right)^{-2} \, ,
\end{equation} 
where the second equality  presumes the relation between $\dm{1}$ and $L$ in Eq.~\ref{eqn:dm1_180Mpc}. We thus see that for neutrino luminosities of $\mathcal{O}(10^{42}) \, \text{erg} \, \text{s}^{-1}$, only sources with $z \lesssim 0.042/\sqrt{\mu}$ can produce $\mu$ltiplets. This is equivalent to the information contained in the left panel of Fig.~\ref{fig:ZvsLum}. Furthermore, for neutrino luminosities such that $\dm{1}/ (c/H_0) \ll \sqrt{\mu}$
the expansion in $z$ is justified.

For $\dm{1} \ll (c/H_0) \sqrt{\mu}$, Eq.~\ref{eqn:multiplet1} simplifies considerably
\begin{eqnarray}
N_{m \geq \mu}^{z \leq z_c} = \nzero \Delta \Omega \, \dm{1}^3 \frac{ \Gamma(\mu) + \lambda_c^{3/2} \Gamma\left(\mu - 3/2, \lambda_c\right) - \Gamma(\mu, \lambda_c)}{3 \lambda_c^{3/2} \Gamma(\mu)} \, . 
\label{eqn:multiplet2}
\end{eqnarray}
Here, we have defined $\lambda_c = [\dm{1}/(z_c c/H_0)]^2$. Specifically, for $z_c \to \infty$,
\begin{align}
N_{m \geq \mu} &\simeq \nzero \Delta \Omega \, \dm{1}^3 \frac{ \Gamma\left(\mu - 3/2\right)}{3 \Gamma(\mu)} \, .
\label{eqn:multiplet3}
\end{align}
which with $\mu = 2$, gives $\nzero \Delta \Omega \, \dm{1}^3 \sqrt{\pi}/3$ as obtained also in Ref.~\cite{Murase:2016gly}. Demanding $N_{m \geq \mu} < 1$ then translates into a constraint on the local source density as a function of luminosity,
\begin{equation}
\nzero < \frac{1}{\Delta\Omega\ \dm{1}^{3}} \frac{3 \Gamma(\mu)}{\Gamma\left(\mu - 3/2\right)} \, .
\label{eqn:PSlimit}
\end{equation}
This limit is shown by way of example for $\mu = 23$ as the black line in Fig.~\ref{fig:limits}. We also assumed that $\dm{1} \simeq 180 \, \text{Mpc} \times \sqrt{L_\nu / (10^{42} \, \mathrm{erg\ s}^{-1})}$, cf.\ Sec.~\ref{sec:IC}.

\subsection{Average test--statistics}
\label{sec:analytical2}

We look for correlations between the distribution of $\Nws$ warm spots and a sky map $g(\hat{n})$ of the matter distribution in the nearby Universe, $\hat{n}$ being a unit vector pointing in a certain direction in the sky. An example of such a tracer could be the 2MRS catalogue that we will discuss in Sec.~\ref{sec:2MRS} below. 

We quantify the degree of correlation defining the (unbinned) log--likelihood as a function of the number of neutrino sources in the nearby Universe $N_a$:
\begin{equation}
\log \mathcal{L}(N_a) = \sum_{i=1}^{\Nws} \log \left( \frac{N_a}{\Nws} \mathcal{S}_i + \left(1 - \frac{N_a}{\Nws} \right) \mathcal{B}_i \right) \, , \label{eqn:loglikelihood}
\end{equation}
where
\begin{equation}
\mathcal{S}_i = g(\hat{n}_i) = \frac{1}{\mathcal{N}} \int_0^{z_c} \mathrm{d} \mathcal{V} \, n_{\text{2MRS}}(z, \hat{n}_i ) \quad \text{with} \quad \mathcal{N} = \int \mathrm{d} \hat{n} \, \int_0^{z_c} \mathrm{d} \mathcal{V} \, n_{\text{2MRS}}(z, \hat{n}_i ) \, , \label{eqn:Si}
\end{equation}
is the signal probability density $g(\hat{n})$ (given e.g.\ by a map of 2MRS galaxies with $z < z_{\text{c}} = 0.02$, see bottom panel of Fig.~\ref{fig:2MRSdist}), evaluated at $\hat{n}_i$, the position of the $i$-th warm spot, and similarly for the isotropic background probability density, i.e.
\begin{equation}
\mathcal{B}_i = b(\hat{n}_i) = \frac{1}{4 \pi} \, . \label{eqn:Bi}
\end{equation}
Note that as probability densities, both $g$ and $b$ are normalised to unity, $\int \dd \hat{n} \, g(\hat{n}) = \int \dd \hat{n} \, b(\hat{n}) = 1$. The parameter $N_a$ is the estimate for the number of sources that are drawn from $g(\hat{n})$ and it is the only free parameter.

By defining the test--statistics (TS) as in Appendix~\ref{sec:averageTS}, we find that its average is
\begin{align}
\langle \mathrm{TS} \rangle =& \frac{N_a^2}{\Nws} 16 \pi^2 \sigma^2 \, , \label{eq:aveTS}
\end{align}
with $\sigma^2$ the variance of $g(\hat{n})$.
The variance $\sigma^2$ of $g$ is easily computed from $g$. For the 2MRS sky map out to $z_{\text{c}} = 0.02$ smoothed with the angular resolution for tracks ($1^{\circ}$ standard deviation), we find $\sigma^2 = 1.66 \times 10^{-2}$ (see Sec.~\ref{sec:2MRS}).

\subsection{Warm spots correlation sensitivity}
\label{sec:analytical3}

From Eq.~\ref{eqn:multiplet2} we compute the number $N_a$ of warm spots that correlates with $g(\hat{n})$ as
\begin{equation}
N_a \simeq \nzero f(\mu, \lambda_c) \, , \label{eq:Na}
\end{equation}
with
\begin{equation}
f(\mu, \lambda_c) \equiv \Delta \Omega \, \dm{1}^3 \frac{ \Gamma(\mu) + \lambda_c^{3/2} \Gamma\left(\mu - 3/2, \lambda_c\right) - \Gamma(\mu, \lambda_c)}{3 \lambda_c^{3/2} \Gamma(\mu)} \, .
\label{eq:f_func}
\end{equation}
The total number of warm spots (see Eq.~\ref{eqn:multiplet3}) is
\begin{equation}
\Nws = N_{\text{bkg}} + N_{m \geq \mu} \simeq N_{\text{bkg}} + \nzero h(\mu, \lambda_c) \, , \label{eq:Nws}
\end{equation}
with
\begin{equation}
h(\mu, \lambda_c) \equiv \Delta \Omega \, \dm{1}^3 \frac{ \Gamma\left(\mu - 3/2\right)}{3 \Gamma(\mu)} \, 
\label{eq:h_func}
\end{equation}
and $N_{\text{bkg}}$ the number of background warm spots.

Substituting Eqs.~\ref{eq:Na} and \ref{eq:Nws} into Eq.~\ref{eq:aveTS} and demanding $\langle \mathrm{TS} \rangle < \mathrm{TS}(p)$, one obtains
\begin{equation}
\nzero \lesssim \frac{\mathrm{TS}(p) h(\mu, \lambda_c)}{2 f^2(\mu, \lambda_c) (4 \pi \sigma)^2} \left[ 1 + \sqrt{1 + \frac{2 N_{\text{bkg}}}{h(\mu, \lambda_c)} \frac{2 f^2(\mu, \lambda_c) (4 \pi \sigma)^2 }{\mathrm{TS}(p) h(\mu, \lambda_c) }}\right] \, .
\label{eqn:CorrSens}
\end{equation}
The above expression gives the local number density of neutrino sources as a function of their neutrino luminosity and $\mu$ltiplicity. In Fig.~\ref{fig:limits}, we have shown this correlation sensitivity $\nzero$ as a function of luminosity $L$ by the blue line. Here, we adopted the parameters $N_{\text{bkg}} = 30$, $\mu = 3$, $z_c = 0.02$, $\Delta \Omega  =4 \pi$ and $\sigma^2 = 2 \times 10^{-2}$ by way of example and again assumed $\dm{1} \simeq 180 \, \text{Mpc} \times \sqrt{L_\nu / (10^{42} \, \mathrm{erg\ s}^{-1})}$, cf.\ Sec.~\ref{sec:IC}.

By considering the low and high neutrino luminosity limits (i.e., $\lambda_c \ll \mu$ and $\lambda_c \gg \mu$), we find (see Appendix~\ref{sec:limits} for more details):
\begin{equation}
\nzero 
\lesssim \frac{\mathrm{TS}(p)}{2 (4 \pi \sigma)^2} \frac{1}{\Delta \Omega} \left( \frac{c}{H_0} z_c \right)^{-3} \left\{ \begin{array}{c l}
	\lambda_c^{-3/2} \dfrac{3 \Gamma(\mu)}{\Gamma(\mu - 3/2)} \left(1 + \sqrt{1 + \dfrac{4 N_{\text{bkg}} (4 \pi \sigma)^2}{\mathrm{TS}(p)}}\right) \quad &\text{for} \quad \lambda_c \ll \mu \, , \\
	9 \lambda_c^{3/2} \dfrac{ \Gamma\left(\mu - 3/2\right)}{3 \Gamma(\mu)} 2 & \text{for} \quad \lambda_c \gg \mu \, .
\end{array} \right. \label{eqn:limit1}
\end{equation}
As it will become clear in the following, cf.\ Sec.~\ref{sec:IC}, the local effective density decreases as the neutrino luminosity increases for  $\lambda_c \ll \mu$ and vice versa for $\lambda_c \gg \mu$. The transition between both regimes is taking place around
\begin{equation}
\lambda_c = 2^{-1/3} \left(\dfrac{\Gamma(\mu)}{ \Gamma\left(\mu - 3/2\right)} \right)^{2/3} \left(1 + \sqrt{1 + \dfrac{4 N_{\text{bkg}} (4 \pi \sigma)^2}{\mathrm{TS}(p)}}\right)^{1/3} \, ,
\label{eqn:Lminimum}
\end{equation}
and the minimum bound is approximately
\begin{equation}
\frac{\mathrm{TS}(p)}{2 (4 \pi \sigma)^2} \frac{1}{\Delta\Omega} \left( \frac{c}{H_0} z_c \right)^{-3} \sqrt{2} \left(1 + \sqrt{1 + \dfrac{4 N_{\text{bkg}} (4 \pi \sigma)^2}{\mathrm{TS}(p)}}\right)^{1/2}
\end{equation}

Equation~\ref{eqn:limit1} can be compared to the limit from the non--detection of PS by IceCube~\cite{Murase:2016gly}, Eq.~\ref{eqn:PSlimit},
\begin{equation}
\nzero \lesssim \frac{1}{\Delta \Omega} \, \left( \frac{c}{H_0} z_c \right)^{-3} \lambda_c^{-3/2} \frac{3 \Gamma(\tilde{\mu})}{ \Gamma\left(\tilde{\mu} - 3/2\right)} \, .
\end{equation}
The number of events $\tilde{\mu}$ for the IceCube PS detection is necessarily larger than the number of events  $\mu$ required to make a warm spot. At low luminosities, the correlation limit is thus more stringent by a factor
\begin{equation}
\frac{2 (4 \pi \sigma)^2}{\mathrm{TS}(p)} \dfrac{\Gamma(\tilde{\mu})}{\Gamma(\mu)} \dfrac{\Gamma(\mu - 3/2)}{\Gamma(\tilde{\mu} - 3/2)} \left(1 + \sqrt{1 + \dfrac{4 N_{\text{bkg}} (4 \pi \sigma)^2}{\mathrm{TS}(p)}}\right)^{-1} \, .
\end{equation}
This gives an improvement by a factor $\sim 20$ for $\mu = 3$, $\tilde{\mu} = 23$, $N_{\text{bkg}} = 33$ and $\sigma^2 = 2 \times 10^{-2}$.

We thus conclude that the correlation sensitivity has the potential to improve bounds on the local source density $\nzero$ as a function of muon neutrino luminosity $\Lnu$. Before investigating this with a detailed MC study described in Sec.~\ref{sec:method}, we introduce the 2MRS catalogue.

\section{2MASS Redshift Survey catalogue}
\label{sec:2MRS}

The 2MASS Redshift Survey (2MRS) \cite{Crook:2006sw} catalogs the distribution of galaxies and dark matter in the local Universe out to a mean redshift of z = 0.03. The final catalog contains 45000 galaxies with photometric redshift information. The distribution of the directions of these galaxies in the sky within different redshift slices is highly anisotropic. For z$<0.02$, clear excesses can be seen along the super-galactic plane, as shown in Fig.~\ref{fig:2MRSdist}. Structures such as the Large Magellanic Cloud as well as the Fornax and Hydra clusters are also visible.
\begin{figure}[tbh]
\centering
\includegraphics[scale=1]{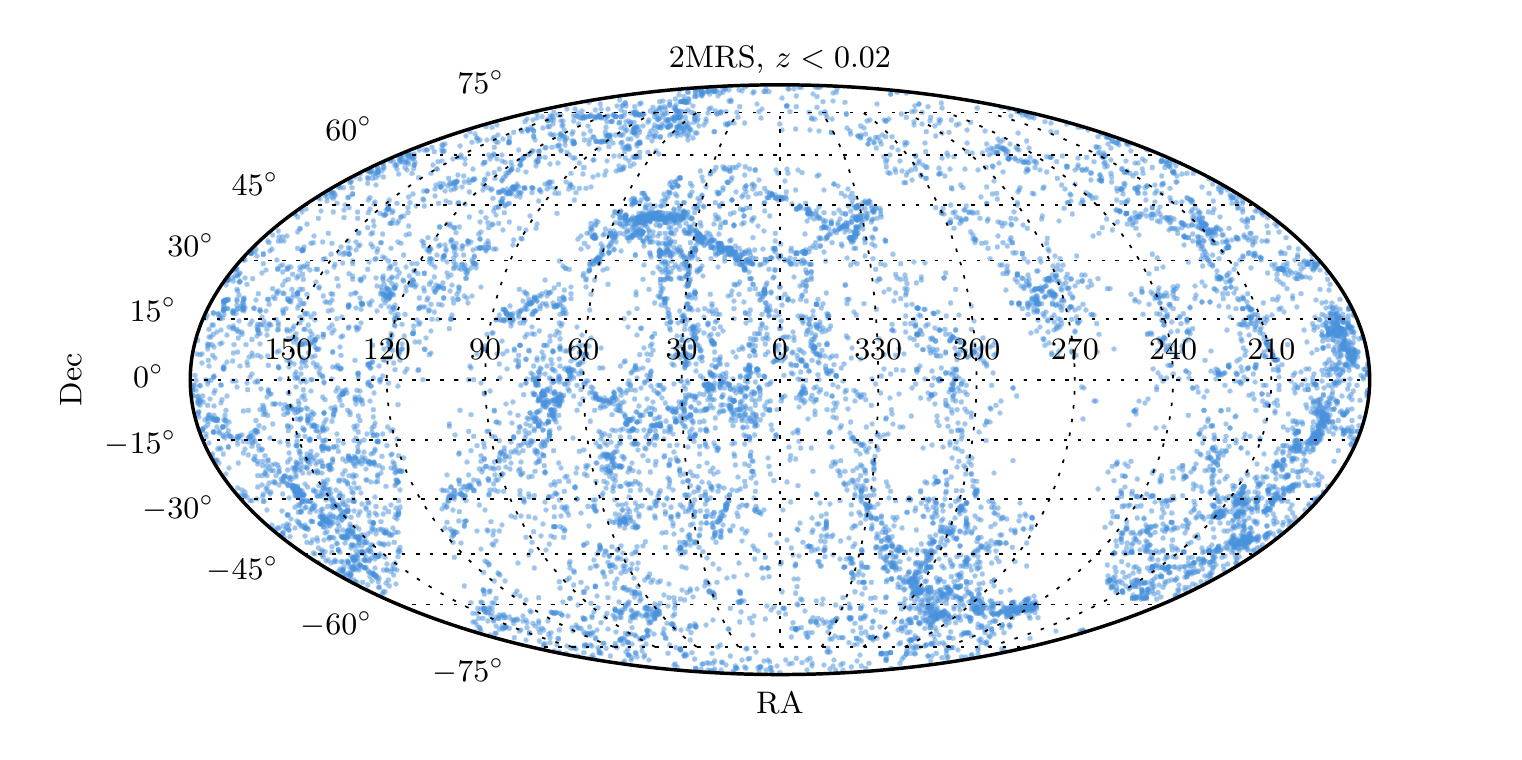}
\includegraphics[scale=1]{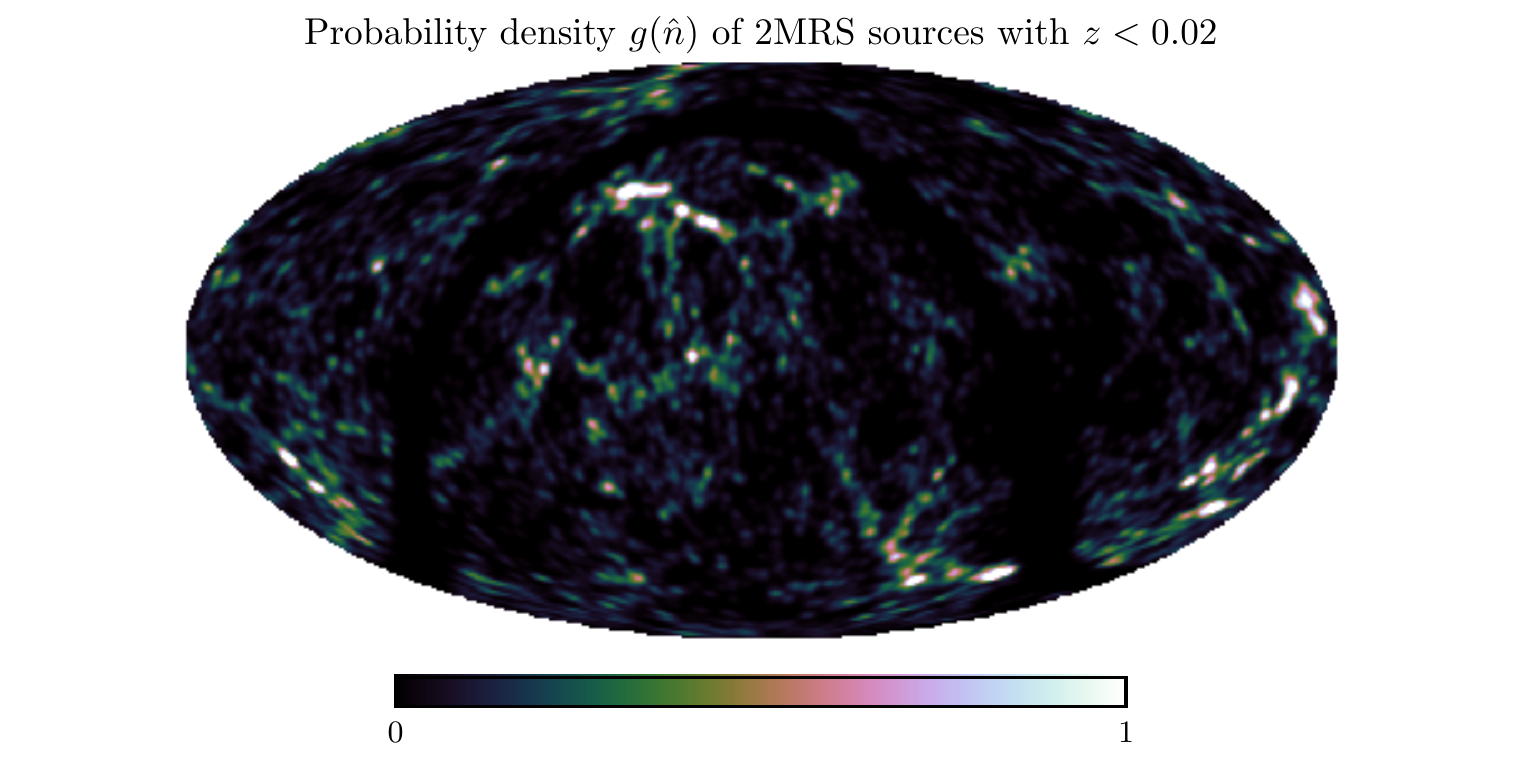}
\caption{\textbf{Top:} Scatter plot of all 2MRS galaxies at redshifts $ z <0.02$. \textbf{Bottom:} Map of all 2MRS galaxies at redshifts $z < 0.02$. This is used as $g(\hat{n})$ in Eq.~\ref{eqn:Si}.}
\label{fig:2MRSdist}
\end{figure}

The 2MRS catalogue could be considered as a good tracer of the anisotropic matter distribution of the local Universe. In fact high-energy neutrinos can be generated via $pp$ or $p\gamma$ interactions within astrophysical sources. In order to produce a sizable amount of neutrinos, protons have to be located in matter dense environments such as active galactic nuclei, blazars, starburst galaxies, gamma-ray bursts, and so  on. These sources are the main constituents of the matter distribution of the local universe constituting the 2MRS catalogue. Assuming that the neutrino warm spots are generated by neutrino events emitted from local astrophysical sources traced by the 2MRS catalogue, in the following we will adopt a source-independent approach and look for correlations of the IceCube warm spots and the 2MRS matter distribution.  

\section{Correlation analysis method}
\label{sec:method}

We explore the sensitivity of a correlation analysis to local effective density of standard candle sources  by using Monte Carlo simulations of the distribution of neutrino PS in our Universe. In this Section, we describe the method adopted in our correlation study.

\subsection{IceCube and IceCube-Gen2 detector properties}
\label{sec:IC}

The normalization $\Phi$ of an $E^{-2}$ astrophysical flux of muon neutrinos required to produce 1 event within the detector at declination $\delta$ is given by
\begin{equation}
\Phi = \frac{1}{\sum_{i=1}^{S}\int_{E_{\mathrm{low}}(\delta)}^{E_{\mathrm{high}}(\delta)}  E^{-2} \mathcal{A}_i(\delta)   dE \times T_i}\ ,
\end{equation}
where the index $i$ runs over the $S$ different data samples, $\mathcal{A}_i(\delta)$ is the effective area of the sample $i$ at declination $\delta$ and $T_i$ is the livetime of the sample $i$. The declination dependent effective areas, livetimes and energy ranges of each sample are obtained from Refs.~\cite{Aartsen:2016oji, Aartsen:2014cva, Abbasi:2010rd}. The energy ranges are approximately $1 \, \text{TeV}$ to $1 \, \text{PeV}$ in the Northern sky and $100 \, \text{TeV}$ to $100 \, \text{PeV}$ in the Southern sky. The median $\dm{1}$ over the whole sky for the seven year search is found to be
\begin{equation}
\dm{1} \sim 180 \, \text{Mpc} \times \sqrt{\frac{L_\nu}{10^{42}{\mathrm{erg\ s}^{-1}}}} \, .
\label{eqn:dm1_180Mpc}
\end{equation}

With IceCube-Gen2 in the Sunflower 240 configuration~\cite{Aartsen:2014njl}, for a PS search with a threshold of $1 \, \text{TeV}$ we can expect $\sim 200$ astrophysical neutrinos per year at the level of the current diffuse flux. From this, we obtain
\begin{equation}
\dm{1} \sim 300 \, \text{Mpc} \times \sqrt{\frac{L_\nu}{10^{42}{\mathrm{erg\ s}^{-1}}}} \, ,
\label{eqn:dm1_300Mpc}
\end{equation}
for a 10 year search. This sample is expected to contain $\sim 15000$ background events per year, which is a factor of $\sim 4$ lower than that of IceCube~\cite{Aartsen:2014njl}.

\subsection{Simulation of the expected IceCube events}
\label{sec:sim}

For a given redshift evolution scenario (see Eq.~\ref{eqn:SrcEvol}),  we generate source redshifts from the expected redshift distribution up to $z_{\mathrm{max}}=3$. For each astrophysical source, if it 
falls beyond the sensitivity threshold distance (see the left panel of Fig. \ref{fig:ZvsLum}), we assume that it contributes only to the diffuse neutrino  flux observed at Earth; otherwise we assign to the source a direction in the sky through random sampling from the 2MRS map of the corresponding redshift slice for $z \le 0.1$ (beyond which the 2MRS catalogue is sparse), or assume that it is isotropically distributed otherwise. While sampling from the 2MRS map, we adopt redshift slices of width 0.01.

Subsequently the neutrino flux expected at Earth for this source is evaluated as described in Sec.~\ref{sec:IC}. If the neutrino flux is high enough  to make a warm spot at a given threshold with significance corresponding to a pre-trial  p-value $p=10^{-3.7}$\cite{Aartsen:2016oji} (approximated to correspond to one third of the flux required for a post trial 3$\sigma$ discovery), then a warm spot is added to the map at the corresponding coordinates and a correlation test is performed according to the unbinned maximum likelihood ratio method described in Eq.~\ref{eqn:loglikelihood}. In each trial, the map also contains a number of isotropically distributed background warm spots, drawn from the expected number of background warm spot distribution~\cite{Aartsen:2016oji}.

This process is repeated until a source with a flux high enough to cause a post trial 3 sigma observation in the PS sky map is obtained. The sensitivity of the warm spot correlation search is the 90th percentile of the density at which the p-value of the warm spot-2MRS correlation crosses 0.5 in many realizations of the Monte Carlo. 
Similarly, the 90\% C.L. upper limit from the non observation of a PS is defined as the 90th percentile of the density just before a post trial 3 sigma observation in the PS sky map is obtained, in many realizations of the Monte Carlo. In total, 1200 trials of the MC were performed.

\section{Results}
\label{sec:results}

In this Section, we present the main results of our correlation analysis for IceCube. We also discuss prospects for IceCube-Gen2.

\subsection{Prospects of neutrino point source detection}

For simplicity we will assume in the following that the sources have an $E_\nu^{-2}$ power law energy spectrum over [1 TeV, 10~PeV]. Figure~\ref{fig:ResultsIceCube} (top panel) shows our results for the IceCube detector. In the plane defined by the local density of neutrino sources and the $\nu_\mu$  luminosity, the upper limit implied by the non--observation of PS in the 7 year search~\cite{Aartsen:2016oji} (``PS limit'') is shown by the solid black line. The maximum allowed local source density $\nzero$ is decreasing as a function of luminosity as $\propto \Lnu^{-3/2}$. This behavior is in agreement with our analytical estimate for $\mu=23$ (cf.\ Sec.~\ref{sec:analytical1}, see also Ref.~\cite{Murase:2016gly}), shown by the dashed black line. $\mu=23$ is the median expected over the whole sky for an $E_\nu^{-2}$ flux at the level required to obtain a post trial 3 sigma discovery. Even though the analytical estimate is derived under a number of simplifying assumptions  and for a sample of benchmark sources, the agreement with the result from the MC study is excellent.

While the scaling with $\Lnu$ had already been discussed in Ref.~\cite{Murase:2016gly}, our overall normalization is different, for example at $L = 10^{39} \, \text{erg} \, \text{s}^{-1}$, our limit is less stringent by a factor of $\sim 30$ (see their Fig.~3). The difference can be attributed to the assumed $\mu$ and the relation between $\dm{1}$ and luminosity $L$. For a low energy sample like the one adopted here, multiplets have been observed and therefore the $\mu$ that can be excluded by PS non--detection is necessarily very large ($\mu \simeq 23$). At the same time, $\dm{1}$ is also relatively large due to the large number of signal events expected in the low energy threshold sample. On the other hand, for a high energy threshold sample as that of Ref.~\cite{Murase:2016gly}, multiplets are less likely and so $\mu=2$ is appropriate. Note, however, that for higher energy thresholds($\geq 200$ TeV) the $\dm{1}$ is also much smaller (by a factor of $\sim 5$). 
If we optimistically assume that no multiplets have been observed in the low energy threshold sample and use the $\dm{1}$ of Eq.~\ref{eqn:dm1_180Mpc}, we reproduce the limits of Ref.~\cite{Murase:2016gly}. 
Although not visible from the plotted $L_\nu$ range, note that we also observe an asymptotical flattening of the black curve in the high luminosity range, similarly to what shown in Fig.~3 of~\cite{Murase:2016gly}.

In Fig.~\ref{fig:ResultsIceCube} (top panel), the blue solid line shows the sensitivity for the local Universe correlation analysis (``correlation sensitivity'') derived in the MC study. The correlation sensitivity curve is lower than the PS limit by almost one order of magnitude at low luminosities, $\Lnu \lesssim 10^{40} \, \text{erg} \, \text{s}^{-1}$. A similar trend has been also discussed in Ref.~\cite{Fang:2016hyv} for the pair method. The maximum allowed density is decreasing $\propto \Lnu^{-3/2}$ for $\Lnu \lesssim 10^{40} \, \text{erg} \, \text{s}^{-1}$ similarly to the black curve of the PS non detection. The correlation sensitivity turns over between $10^{42}$ and $10^{43} \, \text{erg} \, \text{s}^{-1}$ and rises again for larger $\Lnu$. This behaviour is in good agreement with our analytical estimate, cf.\ Eq.~\ref{eqn:CorrSens}, shown by the dashed blue line. The latter has been obtained for $N_{\mathrm{bkg}}=33$, $\mu=3.5$, $z_c =0.02$ and $\sigma^2=1.66 \times 10^{-2}$.

\begin{figure}[htb]
\centering
\includegraphics[scale=0.95]{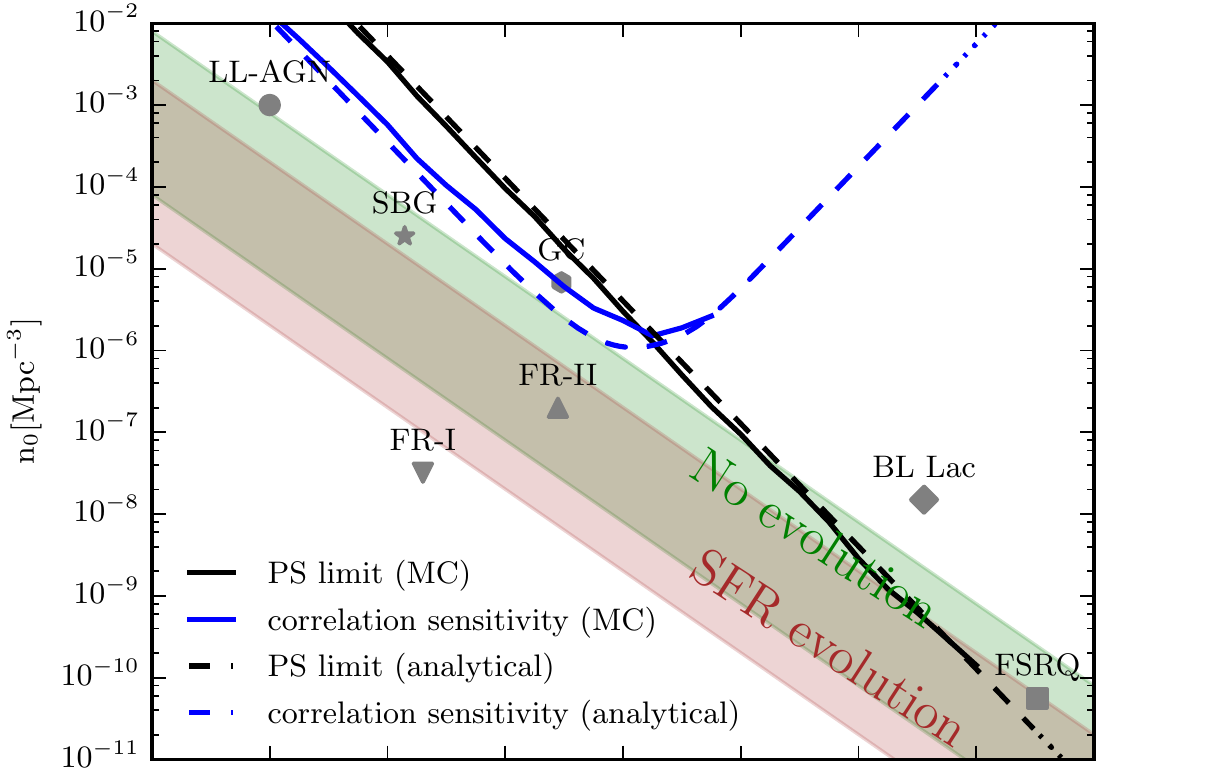}\\
\includegraphics[scale=0.95]{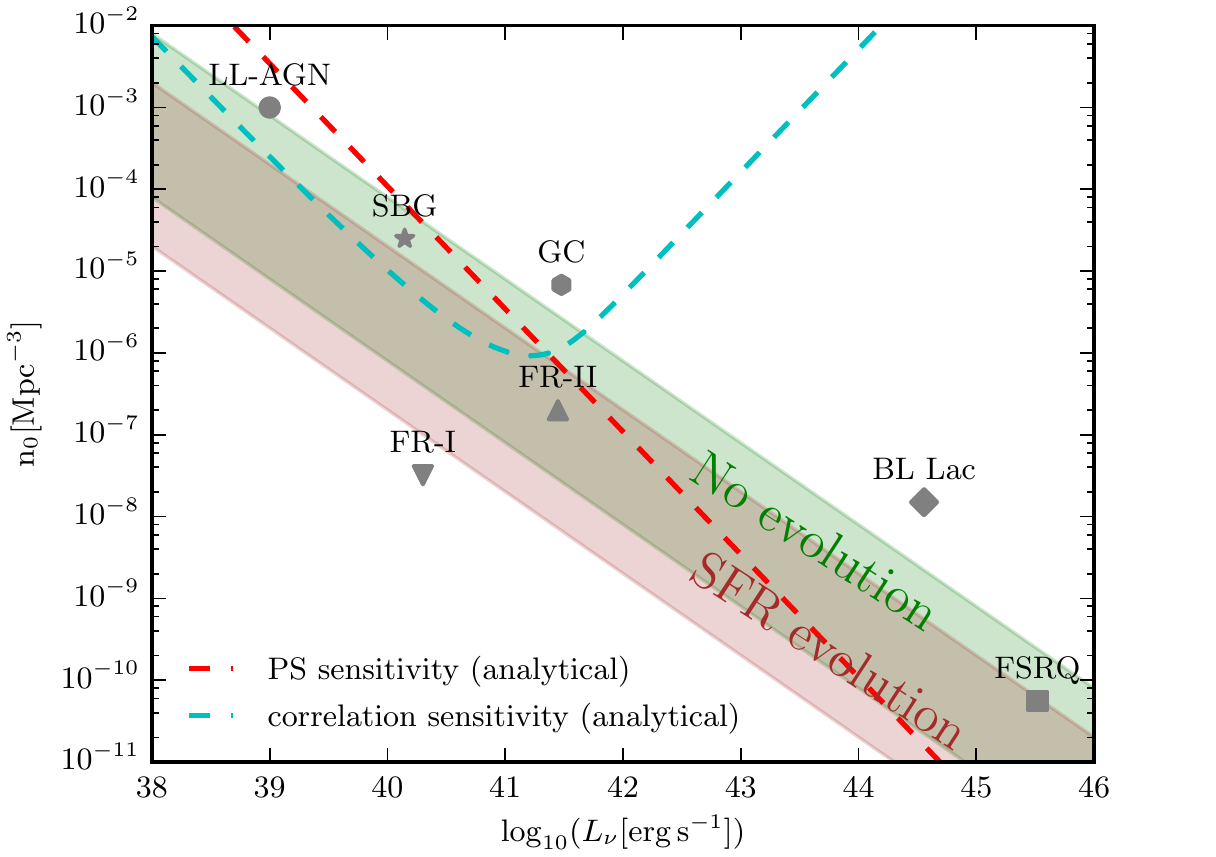}
\caption{Prospects of neutrino PS detection. Top panel: Local density of neutrino sources as a function of $L_\nu$. The solid black line shows the upper limit implied by the non--observation of PS in the 7 year search~\cite{Aartsen:2016oji}, while the dashed black line is its corresponding analytical estimation obtained for $\mu=23$. The solid  blue line represents the sensitivity for the correlation analysis, and the dashed blue line is its corresponding theoretical estimation ($N_{\mathrm{bkg}}=33$, $\mu=3.5$, $z_c =0.02$ and $\sigma^2=1.66 \times 10^{-2}$).  The green and red bands mark the regions of parameter space that would explain the observed diffuse flux~\cite{Aartsen:2015zva} for no-redshift evolution ($q = 0$) and  strong redshift evolution ($q=3$). The top (bottom) of each shaded bands assume that $100 \, \%$ ($1 \, \%$) of the diffuse flux is produced by each $(\nzero, L_\nu)$. The gray markers represent examples of benchmark astrophysical sources (see text for details). Bottom: Same as top panel but the plotted curves have been derived for 10 years of operation of IceCube-Gen 2~\cite{Aartsen:2014njl}. The forecasted value of $\mu$ for non-observation of PS is 7;  while the sensitivity to the correlation analysis has been derived for $N_{\mathrm{bkg}}=33$~\cite{Aartsen:2016oji}, $\mu=2$ and $z_c$ and $\sigma$ as above. For IceCube, the region of the parameter space above the upper limit implied by the non--observation of PS and above the sensitivity curve for the correlation analysis are excluded.}
\label{fig:ResultsIceCube}
\end{figure}

We note that the difference for $10^{40}$ to $10^{42} \, \text{erg} \, \text{s}^{-1}$ between the dashed and solid blue lines is likely due to a number of simplifying assumptions made in the derivation of Eq.~\ref{eqn:CorrSens}: (1) Instead of setting the limit based on the $90 \, \%$ quantile of the distribution of test--statistics, $\text{TS}_{90 \, \%}$, we have worked with the mean test--statistics $\langle \text{TS} \rangle$, which is smaller than $\text{TS}_{90 \, \%}$, making our analytical estimate overly optimistic. (2) In the analytical estimate, we assume that the angular distribution of 2MRS sources is the same at all redshifts $z < 0.02$ whereas it is assumed isotropic beyond. This is not realistic as the angular distribution varies with redshift for $z < 0.02$. This should also render the analytically derived correlation sensitivity too optimistic. (3) In the analytical estimation we neglect any declination dependence of the detector acceptance. 

We further note that our analytical estimate is only valid as long as $\dm{1}/ (c/H_0) \ll \sqrt{\mu}$ which with Eq.~\ref{eqn:dm1_180Mpc} translates into a maximum luminosity $L_\nu = 3.3 \times 10^{45} \, \text{erg} \, \text{s}^{-1}$ for $\mu = 23$ (PS limit) and $5.1 \times 10^{44} \, \text{erg} \, \text{s}^{-1}$ for $\mu = 3.5$ (correlation analysis). Beyond this, the analytical estimate is not valid anymore and we show the extrapolations with dotted lines.

In Fig.~\ref{fig:ResultsIceCube}, we also indicate the regions of parameter space that would explain the observed diffuse flux~\cite{Aartsen:2015zva}. This limit has been obtained by assuming a diffuse neutrino flux $\propto E_\nu^{-2}$, normalized to  $10^{-8}$~GeV~cm$^{-2}$~s$^{-1}$~sr$^{-1}$ and for a redshift evolution as in Ref.~\cite{Yuksel:2008cu} for ``SFR evolution'' or $q =0$ for ``no evolution''. 
The top (bottom) of each shaded bands assume that $100 \, \%$ ($1 \, \%$) of the diffuse flux is produced by the parameter pair $(\nzero, L_\nu)$. These bands have been derived from our MC study and show the expected $L_\nu^{-1}$ scaling.

In Fig.~\ref{fig:ResultsIceCube} (bottom panel), we show analytical estimates for the expected performance of the IceCube-Gen2 detector~\cite{Aartsen:2014njl}. The sensitivity of the all-sky PS search (red dashed curve) is better by a factor of $\sim 40$ with respect to the related IceCube curve due to the lower background expected. As described in Sec.~\ref{sec:IC}, the PS search of IceCube-Gen2 is expected to observe $\sim15000$ background events per year. Consequently, a 10 year IceCube-Gen2 search will have a factor of $~3$ lower background and the $\mu$ that can be excluded by an all sky PS search will also be lower. The resulting smaller number of events $\mu=7$ required to find a PS is the reason for this improvement by a factor of $\sim 40$.

The correlation sensitivity (cyan dashed curve) improves with respect to the IceCube one by a factor of $\sim 50$ at low luminosities. This is also due to the suppression of background such that we have adopted $\mu=2$. The other parameters adopted here are $N_{\mathrm{bkg}}=33$, $z_c =0.02$ and $\sigma^2=1.66 \times 10^{-2}$. We note from Eqs.~\ref{eqn:limit1} and \ref{eqn:Lminimum} that this does not only result in a more stringent expected limit at low $L_\nu$ but also in the minimum of the correlation sensitivity curve moving to lower $L_\nu$ (cf.\ Eq.~\ref{eqn:Lminimum}).

\FloatBarrier

For sources emitting neutrinos of softer spectra with the same integrated luminosity in the $1 \, \text{TeV}$ to $10 \, \text{PeV}$ range, $\dm{1}$ is smaller and consequently both the constraints from PS non detection as well as the sensitivity of the correlation study can be weaker. For $\gamma = 2.6$, the softest unbroken spectrum allowed for sources that contribute to the diffuse flux dominantly~\cite{Aartsen:2015knd}, $\dm{1}$ is lower by a factor of $\sim 3.5$ and both the constraints and sensitivity are worse by a factor of $\sim 50$. For a more reasonable value of $\gamma = 2.2$, favoured by the measurement of the muon neutrino flux from the northern hemisphere~\cite{Aartsen:2016xlq}, $\dm{1}$ is $\sim$34\% smaller and the constraints and the sensitivity scale up by a factor of $\sim 3.5$.

\subsection{Discussion}

In the following, we discuss the implications for the different benchmark sources shown in gray in Fig.~\ref{fig:ResultsIceCube}. Details on the derivation of $(n_0, L_\nu)$ for each source class are reported in Appendix~\ref{sec:n_vs_l}. Given the large astrophysical uncertainties on the modeling of the neutrino production from these sources, we refrain from drawing firm conclusions about the detection prospects from each source class and only aim at providing a qualitative discussion. 

\begin{itemize}

\item Flat-spectrum radio quasars (FSRQ): Those sources are consistent with the diffuse flux limits, however they are marginally disfavored by the IceCube PS limit. They will be finally tested by IceCube-Gen2. We note that the correlation with anisotropies of the local Universe is not helpful in this case as those sources are very bright but have a low local density. Consequently, the chance for them to occur close enough to lead to a significant correlation is very small.

\item BL Lacs: These candidates are already strongly disfavored by the IceCube PS limit. We conclude that they will not be detectable in neutrinos unless our density or luminosity estimates are wrong by more than an order of magnitude. Their $(L_\nu,n_0)$ is not in agreement with the diffuse flux either.

\item Fanaroff-Riley galaxies of type II (FR-II): They are consistent with the diffuse flux band within the astrophysical uncertainties. FR-II are currently not constrained by the upper limit on the non-detection of PS placed by IceCube. The correlation sensitivity analysis of IceCube will not be sensitive to FR-II. They may however be within reach by the IceCube-Gen2 PS searches and correlation analysis when modeling uncertainties are taken into account. 

\item Fanaroff-Riley galaxies of type I (FR-I): Given its low effective neutrino luminosity and local density, this class of sources is well below  the IceCube PS limit and correlation sensitivity curve; FR-I seem to be extremely challenging to detect with available or planned experimental setups. They may become testable with IceCube-Gen2 PS searches only if models of their neutrino emission have been underestimating their luminosity by more than two orders of magnitude.

\item Clusters of Galaxies (GC): They are marginally excluded by the current IceCube PS upper limit and can be further tested by the correlation analysis. Galaxy clusters have not been detected in gamma-rays yet; hence, we will ultimately test the currently adopted mechanism behind the neutrino production with IceCube-Gen2 by employing  all-sky PS searches. Correlation studies performed by IceCube can also test this source class. Galaxy Clusters, being highly extended and well catalogued candidates are however best constrained by targeted stacking searches such as the one described in Ref.~\cite{Aartsen:2014cva}.

\item Starburst galaxies (SBG): Although not very bright in neutrinos with respect to other sources, SBG are very abundant and consistent with the diffuse neutrino flux bands. IceCube-Gen2 standard point source searches will not be sensitive enough to probe their $(L_\nu,n_0)$, but correlation sensitivity searches will ultimately test the current neutrino production model adopted for SBG. 

\item Low--luminosity AGNs (LL-AGN): Although not testable with IceCube, IceCube-Gen2 correlation studies will be able to place limits on the abundance and neutrino luminosity of these sources.

\end{itemize}

Figure~\ref{fig:ResultsIceCube} shows that our correlation analysis can exclude a larger region of the ($L_\nu$,$n_0$) parameter space than the projected PS limit for $L_\nu \lesssim 10^{42}$~erg/s for IceCube  ($L_\nu \lesssim 10^{41}$~erg/s for IceCube-Gen2) and $n_0 \ge 10^{-6}$~Mpc$^{-3}$. We note that the minimum of  the correlation curve moves of one order of magnitude  towards lower neutrino luminosities with  IceCube-Gen 2. 

Besides the assumptions discussed in the previous section on the derivation of the analytical curves with respect to the ones obtained via MC simulations, we also note that we assumed that neutrinos are emitted with an energy spectrum $\propto E_\nu^{-2}$. In principle, the analytical correlation sensitivity is independent of the assumed spectral shape. However, when we express our results as a function of $\Lnu$, the spectral shape enters through Eqs.~\ref{eqn:dm1_180Mpc} or \ref{eqn:dm1_300Mpc} as well as the assumed values of $N_{\text{bkg}}$ and $\mu$. Similarly for the MC analysis, the assumption of an $E_\nu^{-2}$ spectrum enters through the sensitivities (see. Sec.~\ref{sec:sim}). Also, to compare to our benchmark sources, we assumed the $E_\nu^{-2}$ spectral shape. In this sense, tighter limits may be placed by assuming a shallower neutrino energy spectrum.

In this work we mostly focused on steady astrophysical sources. We note as our method can also be applied to transient sources. However, more stringent limits can be obtained for the latter  by focussing on a restricted region of the sky and taking advantage of multi-messenger constraints. 

About $\mathcal{O}(10^3)$~years of operation of IceCube-Gen2 would be needed to probe the whole ($L_\nu$,$n_0$) region of the parameter space where the gray markers of astrophysical sources are sitting via all sky PS searches, especially given the unfavorable ($L_\nu$,$n_0$) of FR-I.  This suggests that targeted searches employing astronomical catalogues may guarantee a more efficient search for neutrinos coming from  certain classes of sources such as FR-I.

\section{Conclusions}
\label{sec:conclusions}

The detection of neutrinos with energy up to few PeV by the IceCube neutrino telescope has ushered in begin of high-energy neutrino astronomy.
Neutrinos of those energies are emitted by yet to be identified cosmic accelerators and could shed light on the origin of cosmic rays and illuminate the physical processes at work in the most violent environments in the Universe.

A good deal of work has been carried out to characterise the high-energy neutrino flux observed by IceCube and pinpoint the sources. Part of all IceCube analyses aims at identifying the sources responsible for the detected diffuse neutrino background and better characterising spectral features. Other searches are instead targeted on neutrino point sources. 

Up to now, dedicated searches looking for neutrino point sources have only placed upper limits. In this work, we exploit the anisotropies of the local Universe to investigate the detection prospects of neutrino steady point sources with IceCube and IceCube-Gen2. Adopting the 2MASS Redshift Survey (2MRS) catalogue as tracer of the anisotropic matter distribution of the local Universe ($z \lesssim  0.02$), we determine analytical estimates for bounds on the sources density as a function of luminosity and perform Monte Carlo simulations of the expected number of neutrino multiplets coming from cosmic accelerators that follow the 2MRS source distribution. 

In order to investigate our future chances of detecting steady point sources via all sky scan searches, we define a plane marked by the neutrino luminosity and local source density ($L_\nu$,$n_0$). We find that for a local density $n_0 \ge 10^{-6}$~Mpc$^{-3}$ and a neutrino luminosity  $L_\nu \lesssim 10^{42}$~erg/s ($L_\nu \lesssim 10^{41}$~erg/s), IceCube (IceCube-Gen2) could provide more stringent limits on the excluded parameter space by adopting a correlation analysis of the neutrino multiplets with the local matter distribution. In this region, a correlation analysis would probe a ($L_\nu$,$n_0$)  parameter space larger than the one constrained by the upper limit of non-detection of point sources.

Sources expected to have $n_0 \le 10^{-6}$~Mpc$^{-3}$ and $L_\nu \le 10^{42}$~erg/s will be difficult to detect even with IceCube-Gen2 unless dedicated stacking searches or searches employing multi-messenger catalogues are performed. For $n_0 \le 10^{-6}$~Mpc$^{-3}$ and $L_\nu \ge 10^{42}$~erg/s the all-sky PS searches will eventually provide the most stringent limits and IceCube-Gen 2 will guarantee a further exploration of the ($n_0$,$L_\nu$) plane with an improvement of one order of magnitude in $L_\nu$.

\section*{Acknowledgements}

We thank Chad Finley, Nora Linn Strotjohann, Kohta Murase and Fabio Zandanel for most helpful discussions and Asen Christov for assistance with the analysis code. This work was supported by the Danish National Research Foundation (DNRF91 and Grant no.\ 1041811001), the Knud H{\o}jgaard Foundation and the Villum Foundation (Project No.\ 13164). Some of the results in this paper have been derived using the HEALPix~\cite{Gorski:2004by} package\footnote{\href{http://healpix.sourceforge.net}{http://healpix.sourceforge.net}}.}

\appendix
\section{Derivation of the average test--statistics}\label{sec:averageTS}

In this Section, we derive the average test statistics as from the (unbinned) log--likelihood introduced in Sec.~\ref{sec:analytical2}.
One can define the test--statistics
\begin{equation}
\mathrm{TS} = 2 \log \frac{\mathcal{L}(\tilde{N}_a)}{\mathcal{L}(N_a = 0)} \, , \label{eqn:TS}
\end{equation}
$\tilde{N}_a$ being the $N_a$ that maximises $\mathcal{L}(N_a)$. According to Wilk's theorem~\cite{Wilks:1938dza}, $\mathrm{TS}$ is drawn from a $\chi^2$ distribution $\chi^2(X, n_{\text{dof}})$ 
with one degree of freedom under the background only hypothesis. The $p$--value then is
\begin{equation}
p = \int_{\mathrm{TS}}^{\infty} \mathrm{d} X \chi^2(X, n_{\text{dof}}=1) \, .
\end{equation}

By adopting Eqs.~\ref{eqn:loglikelihood}-\ref{eqn:Bi} and assuming that $N_a$ represents the true number of sources, Eq.~\ref{eqn:TS} simplifies to
\begin{align}
\mathrm{TS} &= 2 \sum_{i=1}^{\Nws} \log \left[ \frac{N_a}{\Nws} \mathcal{S}_i + \left(1 - \frac{N_a}{\Nws} \right) \mathcal{B}_i \right] - 2 \sum_{i=1}^{\Nws} \log \mathcal{B}_i
\\ &= 2 \sum_{i=1}^{\Nws} \log \left[ 4 \pi \frac{N_a}{\Nws} g(\hat{n}_i) + \left(1 - \frac{N_a}{\Nws} \right) \right] \, .
\end{align}

We now compute the average $\mathrm{TS}$ for the case of $N_a$ sources drawn from $g(\hat{n})$ and $(N_{\text{ws}} - N_a)$ isotropically distributed warm spots:
\begin{align}
\langle \mathrm{TS} \rangle =& 2 N_a \int \mathrm{d} \hat{n} \, g(\hat{n}) \log \left[ 4 \pi \frac{N_a}{\Nws} g(\hat{n}) + \left(1 - \frac{N_a}{\Nws} \right) \right] 
\\ &+ \frac{2 (\Nws - N_a)}{4 \pi} \int \mathrm{d} \hat{n} \log \left[ 4 \pi \frac{N_a}{\Nws} g(\hat{n}) + \left(1 - \frac{N_a}{\Nws} \right) \right] 
\\ =& \frac{2 \Nws}{4 \pi} \int \mathrm{d} \hat{n} \left[ 4 \pi \frac{N_a}{\Nws} g(\hat{n}) + \left(1 - \frac{N_a}{\Nws} \right) \right] \log \left[ 4 \pi \frac{N_a}{\Nws} g(\hat{n}) + \left(1 - \frac{N_a}{\Nws} \right) \right] \, .
\end{align}

We know that the number $N_a$ of neutrino sources in the nearby Universe with $m \geq \mu$ must be much smaller than the total number of warm spots $\Nws$. Therefore, we define $4 \pi g(\hat{n}) \equiv 1 + \delta(\hat{n})$ with $\delta(\hat{n}) \geq -1$ and $\int \mathrm{d} \hat{n} \, \delta(\hat{n}) = 0$, and expand in $\delta(\hat{n}) (N_a / \Nws)$,
\begin{align}
\langle \mathrm{TS} \rangle =& \frac{2 \Nws}{4 \pi} \int \mathrm{d} \hat{n} \left( 1 + \frac{N_a}{\Nws} \delta(\hat{n}) \right) \log \left( 1 + \frac{N_a}{\Nws} \delta(\hat{n}) \right) \\
\simeq& \frac{2 \Nws}{4 \pi} \int \mathrm{d} \hat{n} \left( 1 + \frac{N_a}{\Nws} \delta(\hat{n}) \right) \left[ \frac{N_a}{\Nws} \delta(\hat{n}) - \frac{1}{2} \left( \frac{N_a}{\Nws} \delta(\hat{n}) \right)^2 + \mathcal{O}\left( \frac{N_a}{\Nws} \delta \right)^3(\hat{n})) \right] \\
\simeq & \frac{N_a^2}{4 \pi \Nws} \int \mathrm{d} \hat{n} \, \delta^2(\hat{n}) \, ,
\end{align}
which is just $(16 \pi^2 N_a^2 / \Nws)$ times the variance $\sigma^2$ of $g(\hat{n})$,
\begin{align}
\sigma^2 &\equiv \frac{1}{4 \pi} \int \mathrm{d} \hat{n} \left( g(\hat{n}) - \bar{g} \right)^2
= \frac{1}{4 \pi} \int \mathrm{d} \hat{n} \left( g^2(\hat{n}) - \bar{g}^2 \right)
= \frac{1}{(4 \pi)^3} \int \mathrm{d} \hat{n} \left( (1 + \delta(\hat{n}))^2 - 1 \right)
\\ &= \frac{1}{(4 \pi)^3} \int \mathrm{d} \hat{n} \, \delta^2(\hat{n}) \, . \nonumber
\end{align}

\section{Local source density in the low and high neutrino luminosity regimes}\label{sec:limits}

In the following, we will consider the low and high luminosity limits, corresponding to $(\lambda_c \ll \mu)$ and $(\lambda_c \gg \mu)$ of the expressions introduced in Sec.~\ref{sec:analytical3}. For the function $f$, defined in Eq.~\ref{eq:f_func}, we find
\begin{equation}
f(\mu, \lambda_c) = \Delta \Omega \left( \frac{c}{H_0} z_c \right)^{3} \left\{ \begin{array}{c l}
\lambda_c^{3/2} \dfrac{\Gamma(\mu - 3/2)}{3 \Gamma(\mu)} = h \quad &\text{for} \quad \lambda_c \ll \mu \, , \\
\dfrac{1}{3} & \text{for} \quad \lambda_c \gg \mu \, .
\end{array} \right.
\end{equation}
Note that $N_a = \nzero f(\mu, \lambda_c) \stackrel{\lambda_c \to \infty}{\rightarrow} \nzero \Delta \Omega / 3 \left( (c/H_0) z_c \right)^{3}$.

Similarly, $h$ (Eq.~\ref{eq:f_func}) expressed in $\lambda_c$ is just
\begin{equation}
h(\mu, \lambda_c) = \Delta \Omega \, \left( \frac{c}{H_0} z_c \right)^{3} \lambda_c^{3/2} \frac{ \Gamma\left(\mu - 3/2\right)}{3 \Gamma(\mu)} \, .
\end{equation}
So,
\begin{equation}
\frac{\langle \text{TS} \rangle h(\mu, \lambda_c)}{2 f^2(\mu, \lambda_c) (4 \pi \sigma)^2} 
= \frac{\langle \text{TS} \rangle}{2 (4 \pi \sigma)^2} \frac{1}{\Delta \Omega} \left( \frac{c}{H_0} z_c \right)^{-3} \left\{ \begin{array}{c l}
	\lambda_c^{-3/2} \dfrac{3 \Gamma(\mu)}{\Gamma(\mu - 3/2)} \quad &\text{for} \quad \lambda_c \ll \mu \, , \\
	9 \lambda_c^{3/2} \dfrac{ \Gamma\left(\mu - 3/2\right)}{3 \Gamma(\mu)} & \text{for} \quad \lambda_c \gg \mu \, .
\end{array} \right.
\end{equation}
The factor in parentheses in Eq.~\ref{eqn:CorrSens} gives,
\begin{equation}
\left( 1 + \sqrt{1 + \frac{2 N_{\text{bkg}}}{h(\mu, \lambda_c)} \frac{2 f^2(\mu, \lambda_c) (4 \pi \sigma)^2 }{\langle \text{TS} \rangle h(\mu, \lambda_c) }}\right) = \left\{ \begin{array}{c l}
1 + \sqrt{1 + \dfrac{4 N_{\text{bkg}} (4 \pi \sigma)^2}{\langle \text{TS} \rangle}} \quad &\text{for} \quad \lambda_c \ll \mu \, , \\
2 & \text{for} \quad \lambda_c \gg \mu \, ,
\end{array} \right.
\end{equation}
and putting everything together we find
\begin{equation}
\nzero 
= \frac{\langle \text{TS} \rangle}{2 (4 \pi \sigma)^2} \frac{1}{\Delta \Omega} \left( \frac{c}{H_0} z_c \right)^{-3} \left\{ \begin{array}{c l}
	\lambda_c^{-3/2} \dfrac{3 \Gamma(\mu)}{\Gamma(\mu - 3/2)} \left(1 + \sqrt{1 + \dfrac{4 N_{\text{bkg}} (4 \pi \sigma)^2}{\langle \text{TS} \rangle}}\right) \quad &\text{for} \quad \lambda_c \ll \mu \, , \\
	9 \lambda_c^{3/2} \dfrac{ \Gamma\left(\mu - 3/2\right)}{3 \Gamma(\mu)} 2 & \text{for} \quad \lambda_c \gg \mu \, .
\end{array} \right. \label{eqn:limit}
\end{equation}

\section{Neutrino luminosity and local density of steady astrophysical sources}
\label{sec:n_vs_l}

In this Appendix, we briefly report the estimates of the local density and luminosity for the steady sources reported in Fig.~\ref{fig:ResultsIceCube}. In general, not all sources of one class need to be the same, but they can vary in spectral neutrino intensity $I_\nu(E)$ and in overall luminosity $L_\nu$, forming a luminosity distribution. However, it helps to think about ``benchmark'' sources, that is sources of a certain class with the same luminosity. 

If $L^2 (\dd n/ \dd L)$, where $\dd n/ \dd L$ is the luminosity distribution e.g.\ in gamma--rays, is peaked at a certain luminosity, the overall flux will be dominated by sources within a narrow luminosity range around this peak luminosity. For instance if $\dd n/ \dd L \propto L^{\alpha}$ with $\alpha > -2$ below and  $\alpha < -2$ below and above a break luminosity, the break luminosity can be used as the benchmark luminosity. Specifically, Ref.~\cite{Murase:2016gly} choses the peak neutrino luminosity to be the luminosity $L_{\nu}$ that maximises $\Lnu \dd n / \dd (\ln \Lnu)$
\begin{equation}
L_{\nu} = \argmax\limits_{\tilde{L}_{\nu}} \tilde{L}_{\nu} \frac{\dd n}{\dd \ln L_{\gamma}} \, ,
\end{equation}
and then defines an effective local source density by requiring that the benchmark luminosity matches the integrated luminosity of the distribution,
\begin{equation}
n_0  = \frac{1}{L_{\nu}} \int \dd (\ln L_{\gamma}) \Lnu \frac{\dd n}{\dd \ln L_{\gamma}} \, . \label{eqn:defn0eff}
\end{equation}
Note that this requires to define the scaling between $L_{\nu}$ and, e.g., $L_{\gamma}$ only up to a constant, e.g.\ $L_{\nu} \propto (L_{\gamma})^{\eta}$. The normalisation can then be determined by an additional assumption, e.g.\ that the diffuse neutrino flux be matched.

The link between the neutrino and the gamma-ray luminosity is also established defined by studying the $pp$ or $p\gamma$ interactions driving the production of neutrinos and gamma-rays in the source. In fact if sources happen to be observed, e.g., in gamma--rays, then we can use scaling relations to relate the observed gamma--ray and the expected neutrino luminosities. Astrophysical high-energy neutrinos are usually produced by the decay of charged pions from hadronic interactions of cosmic rays with matter (proton-proton interactions, $pp$) and radiation (proton-photon interactions, $p\gamma$). The same mechanisms also guarantee  the production of $\gamma$-rays from neutral pions. The intensities of each neutrino flavor and gamma-rays are related (see, e.g., Ref.~\cite{Ahlers:2014ioa})
\begin{equation}
 I_\nu(E_\nu) \simeq K I_\gamma (E_\gamma)\ , \label{eqn:GammaNuScaling}
\end{equation} 
with $E_\gamma = 2 E_\nu$ and $K \simeq 2\ (K \simeq 1)$  for $pp$ ($p\gamma$) interactions. The density of those sources is usually assumed to be the same as estimated in other wavelengths. By employing the relations above and assuming an injection spectral index $\Gamma = 2$ for all sources emitting neutrinos, we now estimate the local density and corresponding neutrino effective luminosity for the following sources:

\begin{itemize}

\item {\bf Starburst galaxies.} 
Neutrinos are produced in starburst galaxies through $pp$ interactions, see e.g.~Ref.~\cite{Waxman:2015ues}. The infrared (IR) luminosity  of these sources~\cite{Gruppioni:2013jna}, measured in the range $[8, 10^3]\ \mu$m, is linearly related to the corresponding $\gamma$-ray luminosity observed by {\it Fermi} in the [0.1,100]~GeV~\cite{Ackermann:2012vca}: $\log(L_{\gamma}/\mathrm{erg\ s}^{-1}) = \alpha \log(L_{\mathrm{IR}}/{10^{10} L_\odot})+ \beta$ with $L_\odot$ the solar luminosity, $\alpha = 1.17 \pm 0.07$ and $\beta = 39.28 \pm 0.08$. By assuming a break luminosity in the IR luminosity function as $L_{\mathrm{IR}} \simeq 10^{11} L_\odot$ from~\cite{Gruppioni:2013jna}, the corresponding gamma-ray luminosity is $L_{\gamma} = 2.8 \times 10^{40}$~erg/s\footnote{Note as our estimations are conservative  due to a possibly underestimated contribution of the ultra luminous infrared galaxies, see e.g.~Refs.~\cite{Peng:2016nsx,Griffin:2016wzb}.}. The latter is proportional to the neutrino luminosity via  eq.~\ref{eqn:GammaNuScaling}, $L_{\nu} = 1.4 \times 10^{40}$~erg/s.  The local density of starburst galaxies is estimated to be $n_0 \simeq 2.5 \times 10^{-5}$~Mpc$^{-3}$~\cite{Gruppioni:2013jna}.  

\item {\bf Cluster of galaxies.}
Neutrinos in galaxy clusters are also produced through $pp$ interactions, see e.g.~\cite{Zandanel:2014pva}. Given the uncertainties in the modeling of the neutrino production in clusters, we here rely on the phenomenological luminosity-mass relation. Following Table~3 of~\cite{Zandanel:2014pva}, we extrapolate the expected neutrino luminosity of five typical clusters:  $L_{\nu, {\mathrm{Perseus}}} \simeq 4.7 \times 10^{41}$ erg/s, $L_{\nu, {\mathrm{Virgo}}} \simeq 3.3 \times 10^{41}$ erg/s, $L_{\nu, {\mathrm{Centaurus}}} \simeq 7.6 \times 10^{40}$ erg/s, $L_{\nu, {\mathrm{Coma}}} \simeq 1.2 \times 10^{42}$ erg/s , and $L_{\nu, {\mathrm{Ophiuchus}}} \simeq 2.9 \times 10^{42}$ erg/s.  We then assume that our benchmark neutrino luminosity for this source class is given by the average of the above luminosities: $L_{\nu} \simeq  10^{42}$~erg/s. The local density of galaxy clusters with masses above $9 \times 10^{13}\ M_\odot$ is equal to $n_0 \simeq 6.8 \times 10^{-6}$~Mpc$^{-3}$~\cite{Zandanel:2014pva}. 

\item \textbf{Flat--spectrum radio quasars (FSRQs).}
Neutrinos can be produced by p$\gamma$ interactions in FSRQs~\cite{Murase:2015ndr}. We adopt the local gamma--ray luminosity function as measured by \textit{Fermi}~\cite{Ajello:2011zi}, $dn/dL_{\gamma} = A/ [\left({L_{\gamma}}/{L_{\star}}\right)^{\gamma_1} + \left({L_{\gamma}}/{L_{\star}}\right)^{\gamma_2}]$, with $L_{\gamma}$ defined in the $[0.1, 100]$~GeV band. As $\gamma_1 < 2$ and $\gamma_2 > 2$, the luminosity in logarithmic luminosity intervals, $L_{\gamma} \, \dd n/\dd (\ln L_{\gamma}) = L_{\gamma}^2 (\dd n / \dd L_{\gamma})$ has a local maximum close to $L_{\star} = 2.2 \times 10^{47} \, \text{erg} \, \text{s}^{-1}$, which we find to be at $L^{\text{eff}}_{\gamma} \equiv 9.2 \times 10^{46} \, \text{erg} \, \text{s}^{-1}$. The overall luminosity is thus dominated by a small luminosity interval around $L^{\text{eff}}_{\gamma}$. We follow Ref.~\cite{Murase:2016gly} in defining the corresponding local source density $n_0$ through Eq.~\ref{eqn:defn0eff}. We find $n_0 \simeq 5.6 \times 10^{-11} \, \text{Mpc}^{-3}$ which should be compared to the total local number density of FSQRs, $n_{0}^\prime \simeq 5 \times  10^{-10} \, \text{Mpc}^{-3}$~\cite{Ajello:2013lka}. Scaling the gamma--ray luminosity via Eq.~\ref{eqn:GammaNuScaling} gives $L_{\nu} \simeq L_{\gamma}/ 4 / \ln(1000) \simeq 3.3 \times 10^{45} \, \text{erg} \, \text{s}^{-1}$ where the factor $\ln(1000)$ comes from converting from $[0.1, 100]$~GeV luminosity to unit logarithmic energy interval luminosity.

\item \textbf{BL Lacs.} 
Neutrinos are produced by p$\gamma$ interactions in BL Lacs, e.g.~\cite{Murase:2015ndr}. We adopt a gamma--ray luminosity function that should approximate the local one as reported by \textit{Fermi}~\cite{Ajello:2013lka}, $\Phi(L_{\gamma}) = A/ [\left({L_{\gamma}}/{L_{\star}}\right)^{\gamma_1} + \left({L_{\gamma}}/{L_{\star}}\right)^{\gamma_2}]$, with $A= 3 \times 10^{-7} \, \text{Mpc}^{-3} \, \text{erg}^{-1} \, \text{s}$, $L_{\star} = 10^{46} \, \text{erg} \, \text{s}^{-1}, \gamma_1 = 2$ and $\gamma_2 = 3.5$. Unlike in the case of FSRQs, $L_{\gamma}^2 (\dd n / \dd L_{\gamma})$ does not have a local maximum and therefore the overall luminosity is not dominated by a narrow luminosity interval. We nevertheless adopt the break luminosity $L_{\star} = 10^{46} \, \text{erg} \, \text{s}^{-1}$ as the benchmark luminosity $L_{\gamma}$, and determine the corresponding local source density $n_0$ through eq.~\ref{eqn:defn0eff}. We find $n_0 \simeq 1.5 \times 10^{-8} \, \text{Mpc}^{-3}$ which is to be compared to the total local number density of BL Lacs, $n_0^\prime \simeq 2 \times10^{-7} \, \text{Mpc}^{-3}$~\cite{Ajello:2013lka}. Scaling the gamma--ray luminosity via Eq.~\ref{eqn:GammaNuScaling} gives $L_{\nu} \simeq L_{\gamma}/ 4 / \ln(1000) \simeq 3.6 \times 10^{44} \, \text{erg} \, \text{s}^{-1}$. 

\item  \textbf{Fanaroff-Riley galaxies  (FR-I and FR-II).} 
Ref.~\cite{Inoue:2011bm} explored the correlation between the {\it Fermi} gamma-ray luminosity and the radio-loud luminosity function of radio galaxies: $\log (L_\gamma) = (-3.90\pm0.61) + (1.16\pm 0.02) \log (L_{5\ {\mathrm{GHz}}})$ for $L_\gamma \in [0.1,10]$~GeV. Depending on the jet morphology, we distinguish between  Faranoff-Riley galaxies  of type I (FR-I) and  type II (FR-II). We approximate the luminosity of those sources to the one corresponding  the break of their luminosity functions defined in Ref.~\cite{Inoue:2011bm}:  $L_{\mathrm{radio},\mathrm{FR-I}} = 10^{26.48}$~W Hz$^{-1}$ sr$^{-1}$ equivalent to  $L_{\gamma,\mathrm{FR-I}} = 4.6 \times 10^{41}$~erg/s and $L_{\mathrm{radio},\mathrm{FR-II}} = 10^{27.39}$~W Hz$^{-1}$ sr$^{-1}$ equivalent to $L_{\gamma,\mathrm{FR-II}} = 5.22 \times 10^{42}$~erg/s. The neutrino luminosity is $L_{\nu,\mathrm{FR-I}} \simeq L_{\gamma,\mathrm{FR-I}}/2 \simeq 5 \times 10^{40}$~erg/s assuming that neutrinos are produced via $pp$ interactions (see e.g.~Ref.~\cite{Tjus:2014dna}) and $L_{\nu,\mathrm{FR-II}} \simeq L_{\gamma,\mathrm{FR-II}}/4 \simeq 2.8 \times 10^{41}$~erg/s if the neutrino production is dominated by $p\gamma$ interactions (e.g., Ref.~\cite{Becker:2005ya}). The local density of these sources is $n_{0,\mathrm{FR-I}} \simeq 3.2 \times 10^{-8}$~Mpc$^{-3}$ and   $n_{0,\mathrm{FR-II}} \simeq 2 \times 10^{-7}$~Mpc$^{-3}$~\cite{Inoue:2011bm}.

\item  \textbf{Low-luminosity active galactic nuclei (LL-AGN).} We assume the benchmark neutrino luminosity $L_{\nu}\simeq 7.8 \times 10^{38}$~erg/s as from the reference model  of Ref.~~\cite{Kimura:2014jba} derived by adopting the H$\alpha$ luminosity function dominated by $L_{H\alpha} \simeq 10^{40}$~erg/s under the assumption of no redshift evolution. The corresponding local density is $n_0^\prime \simeq 1.3 \times 10^{-2}$~Mpc$^{-3}$ corresponding to an effective density $n_0 \simeq 10^{-3}$~Mpc$^{-3}$~\cite{Murase:2016gly}.

\end{itemize}

We summarise our estimates in Table~\ref{tbl1}.

\begin{table}
\centering
\caption{Effective local density and muon neutrino luminosity of various astrophysical steady sources producing high-energy neutrinos in the same energy range of the IceCube neutrino events adopted in this paper.}
\begin{tabular}{c | c | c}
\hline\hline
Source class	&	  $n_0 [\text{Mpc}^{-3}]$	& $L_{\nu} [\text{erg} \, \text{s}^{-1}]$ \\
\hline\hline
Staburst galaxies	&  $2.5 \times 10^{-5}$ & $1.4 \times 10^{40}$ \\
Clusters of galaxies	& $6.8 \times 10^{-6}$	& $1 \times 10^{42}$		 \\
FSRQs			& $5.6 \times 10^{-11}$	& $3.3 \times 10^{45}$ \\
BL Lacs			& $1.5 \times 10^{-8}$	& $3.6 \times 10^{44}$ \\
FR-I                           &  $3.2 \times 10^{-8}$ & $2 \times 10^{40}$\\
FR-II                           &  $2 \times 10^{-7}$ & $2.8 \times 10^{41}$\\
LL-AGN                        &  $1 \times 10^{-3}$ & $1. \times 10^{39}$\\
\hline
\end{tabular}
\label{tbl1}
\end{table}

\newpage

\bibliographystyle{JHEP}
\bibliography{biblio_ICPS2MRS}

\providecommand{\href}[2]{#2}\begingroup\raggedright\begin{thebibliography}{10}

\bibitem{Aartsen:2013bka}
{\bf IceCube} Collaboration, M.~G. Aartsen et~al., {\it {First observation of
  PeV-energy neutrinos with IceCube}},  { Phys. Rev. Lett.} {\bf 111} (2013)
  021103, [\href{http://arxiv.org/abs/1304.5356}{{\tt arXiv:1304.5356}}].

\bibitem{Aartsen:2013jdh}
{\bf IceCube} Collaboration, M.~G. Aartsen et~al., {\it {Evidence for
  High-Energy Extraterrestrial Neutrinos at the IceCube Detector}},  { Science}
  {\bf 342} (2013) 1242856, [\href{http://arxiv.org/abs/1311.5238}{{\tt
  arXiv:1311.5238}}].

\bibitem{Aartsen:2014gkd}
{\bf IceCube} Collaboration, M.~G. Aartsen et~al., {\it {Observation of
  High-Energy Astrophysical Neutrinos in Three Years of IceCube Data}},  {
  Phys. Rev. Lett.} {\bf 113} (2014) 101101,
  [\href{http://arxiv.org/abs/1405.5303}{{\tt arXiv:1405.5303}}].

\bibitem{Aartsen:2014muf}
{\bf IceCube} Collaboration, M.~G. Aartsen et~al., {\it {Atmospheric and
  astrophysical neutrinos above 1 TeV interacting in IceCube}},  { Phys. Rev.}
  {\bf D91} (2015), no.~2 022001, [\href{http://arxiv.org/abs/1410.1749}{{\tt
  arXiv:1410.1749}}].

\bibitem{Aartsen:2015knd}
{\bf IceCube} Collaboration, M.~G. Aartsen et~al., {\it {A combined
  maximum-likelihood analysis of the high-energy astrophysical neutrino flux
  measured with IceCube}},  { Astrophys. J.} {\bf 809} (2015), no.~1 98,
  [\href{http://arxiv.org/abs/1507.03991}{{\tt arXiv:1507.03991}}].

\bibitem{Aartsen:2015rwa}
{\bf IceCube} Collaboration, M.~G. Aartsen et~al., {\it {Evidence for
  Astrophysical Muon Neutrinos from the Northern Sky with IceCube}},  { Phys.
  Rev. Lett.} {\bf 115} (2015), no.~8 081102,
  [\href{http://arxiv.org/abs/1507.04005}{{\tt arXiv:1507.04005}}].

\bibitem{Waxman:1998yy}
E.~Waxman and J.~N. Bahcall, {\it {High-energy neutrinos from astrophysical
  sources: An Upper bound}},  { Phys. Rev.} {\bf D59} (1999) 023002,
  [\href{http://arxiv.org/abs/hep-ph/9807282}{{\tt hep-ph/9807282}}].

\bibitem{Anchordoqui:2013dnh}
L.~A. Anchordoqui et~al., {\it {Cosmic Neutrino Pevatrons: A Brand New Pathway
  to Astronomy, Astrophysics, and Particle Physics}},  { JHEAp} {\bf 1-2}
  (2014) 1--30, [\href{http://arxiv.org/abs/1312.6587}{{\tt arXiv:1312.6587}}].

\bibitem{Halzen:2013dva}
F.~Halzen, {\it {The highest energy neutrinos: first evidence for cosmic
  origin}},  { Nuovo Cim.} {\bf C037} (2014), no.~03 117--132,
  [\href{http://arxiv.org/abs/1311.6350}{{\tt arXiv:1311.6350}}]. [Astron.
  Nachr.335,507(2014)].

\bibitem{Ahlers:2015lln}
M.~Ahlers and F.~Halzen, {\it {High-energy cosmic neutrino puzzle: a review}},
  { Rept. Prog. Phys.} {\bf 78} (2015), no.~12 126901.

\bibitem{Aartsen:2015zva}
{\bf IceCube} Collaboration, M.~G. Aartsen et~al., {\it {The IceCube Neutrino
  Observatory - Contributions to ICRC 2015 Part II: Atmospheric and
  Astrophysical Diffuse Neutrino Searches of All Flavors}},  in { {Proceedings,
  34th International Cosmic Ray Conference (ICRC 2015): The Hague, The
  Netherlands, July 30-August 6, 2015}}, 2015.
\newblock \href{http://arxiv.org/abs/1510.05223}{{\tt arXiv:1510.05223}}.

\bibitem{Tjus:2015rck}
J.~Becker~Tjus, {\it {Search for Galactic cosmic ray sources: The
  multimessenger approach}},  { EPJ Web Conf.} {\bf 105} (2015) 00003.

\bibitem{Waxman:2015ues}
E.~Waxman, {\it {The origin of IceCube's neutrinos: Cosmic ray accelerators
  embedded in star forming calorimeters}},
  \href{http://arxiv.org/abs/1511.00815}{{\tt arXiv:1511.00815}}.

\bibitem{Murase:2013rfa}
K.~Murase, M.~Ahlers, and B.~C. Lacki, {\it {Testing the Hadronuclear Origin of
  PeV Neutrinos Observed with IceCube}},  { Phys. Rev.} {\bf D88} (2013),
  no.~12 121301, [\href{http://arxiv.org/abs/1306.3417}{{\tt
  arXiv:1306.3417}}].

\bibitem{Tamborra:2014xia}
I.~Tamborra, S.~Ando, and K.~Murase, {\it {Star-forming galaxies as the origin
  of diffuse high-energy backgrounds: Gamma-ray and neutrino connections, and
  implications for starburst history}},  { JCAP} {\bf 1409} (2014) 043,
  [\href{http://arxiv.org/abs/1404.1189}{{\tt arXiv:1404.1189}}].

\bibitem{Ando:2015bva}
S.~Ando, I.~Tamborra, and F.~Zandanel, {\it {Tomographic Constraints on
  High-Energy Neutrinos of Hadronuclear Origin}},  { Phys. Rev. Lett.} {\bf
  115} (2015), no.~22 221101, [\href{http://arxiv.org/abs/1509.02444}{{\tt
  arXiv:1509.02444}}].

\bibitem{Senno:2015tra}
N.~Senno, P.~M\'esz\'aros, K.~Murase, P.~Baerwald, and M.~J. Rees, {\it
  {Extragalactic star-forming galaxies with hypernovae and supernovae as
  high-energy neutrino and gamma-ray sources: the case of the 10 TeV neutrino
  data}},  { Astrophys. J.} {\bf 806} (2015), no.~1 24,
  [\href{http://arxiv.org/abs/1501.04934}{{\tt arXiv:1501.04934}}].

\bibitem{Chakraborty:2015sta}
S.~Chakraborty and I.~Izaguirre, {\it {Diffuse neutrinos from extragalactic
  supernova remnants: Dominating the 100 TeV IceCube flux}},  { Phys. Lett.}
  {\bf B745} (2015) 35--39, [\href{http://arxiv.org/abs/1501.02615}{{\tt
  arXiv:1501.02615}}].

\bibitem{Chang:2014sua}
X.-C. Chang, R.-Y. Liu, and X.-Y. Wang, {\it {Star-forming galaxies as the
  origin of the IceCube PeV neutrinos}},  { Astrophys. J.} {\bf 805} (2015),
  no.~2 95, [\href{http://arxiv.org/abs/1412.8361}{{\tt arXiv:1412.8361}}].

\bibitem{Fang:2016amf}
K.~Fang and A.~V. Olinto, {\it {High-energy neutrinos from sources in clusters
  of galaxies}},  { Astrophys. J.} {\bf 828} (2016), no.~1 37,
  [\href{http://arxiv.org/abs/1607.00380}{{\tt arXiv:1607.00380}}].

\bibitem{Zandanel:2014pva}
F.~Zandanel, I.~Tamborra, S.~Gabici, and S.~Ando, {\it {High-energy gamma-ray
  and neutrino backgrounds from clusters of galaxies and radio constraints}},
  { Astron. Astrophys.} {\bf 578} (2015) A32,
  [\href{http://arxiv.org/abs/1410.8697}{{\tt arXiv:1410.8697}}].

\bibitem{Meszaros:2015krr}
P.~M\'esz\'aros, {\it {Gamma Ray Bursts as Neutrino Sources}},
  \href{http://arxiv.org/abs/1511.01396}{{\tt arXiv:1511.01396}}.

\bibitem{Murase:2015ndr}
K.~Murase, {\it {Active Galactic Nuclei as High-Energy Neutrino Sources}},
  \href{http://arxiv.org/abs/1511.01590}{{\tt arXiv:1511.01590}}.

\bibitem{Neronov:2016ksj}
A.~Neronov, D.~V. Semikoz, and K.~Ptitsyna, {\it {Strong constraint on hadronic
  models of blazar activity from Fermi and IceCube stacking analysis}},
  \href{http://arxiv.org/abs/1611.06338}{{\tt arXiv:1611.06338}}.

\bibitem{Padovani:2015mba}
P.~Padovani, M.~Petropoulou, P.~Giommi, and E.~Resconi, {\it {A simplified view
  of blazars: the neutrino background}},  { Mon. Not. Roy. Astron. Soc.} {\bf
  452} (2015), no.~2 1877--1887, [\href{http://arxiv.org/abs/1506.09135}{{\tt
  arXiv:1506.09135}}].

\bibitem{Murase:2013ffa}
K.~Murase and K.~Ioka, {\it {TeV?PeV Neutrinos from Low-Power Gamma-Ray Burst
  Jets inside Stars}},  { Phys. Rev. Lett.} {\bf 111} (2013), no.~12 121102,
  [\href{http://arxiv.org/abs/1306.2274}{{\tt arXiv:1306.2274}}].

\bibitem{Murase:2015xka}
K.~Murase, D.~Guetta, and M.~Ahlers, {\it {Hidden Cosmic-Ray Accelerators as an
  Origin of TeV-PeV Cosmic Neutrinos}},  { Phys. Rev. Lett.} {\bf 116} (2016),
  no.~7 071101, [\href{http://arxiv.org/abs/1509.00805}{{\tt
  arXiv:1509.00805}}].

\bibitem{Senno:2015tsn}
N.~Senno, K.~Murase, and P.~M\'esz\'aros, {\it {Choked Jets and Low-Luminosity
  Gamma-Ray Bursts as Hidden Neutrino Sources}},  { Phys. Rev.} {\bf D93}
  (2016), no.~8 083003, [\href{http://arxiv.org/abs/1512.08513}{{\tt
  arXiv:1512.08513}}].

\bibitem{Tamborra:2015qza}
I.~Tamborra and S.~Ando, {\it {Diffuse emission of high-energy neutrinos from
  gamma-ray burst fireballs}},  { JCAP} {\bf 1509} (2015), no.~09 036,
  [\href{http://arxiv.org/abs/1504.00107}{{\tt arXiv:1504.00107}}].

\bibitem{Tamborra:2015fzv}
I.~Tamborra and S.~Ando, {\it {Inspecting the supernova?gamma-ray-burst
  connection with high-energy neutrinos}},  { Phys. Rev.} {\bf D93} (2016),
  no.~5 053010, [\href{http://arxiv.org/abs/1512.01559}{{\tt
  arXiv:1512.01559}}].

\bibitem{Wang:2015mmh}
X.-Y. Wang and R.-Y. Liu, {\it {Tidal disruption jets of supermassive black
  holes as hidden sources of cosmic rays: explaining the IceCube TeV-PeV
  neutrinos}},  { Phys. Rev.} {\bf D93} (2016), no.~8 083005,
  [\href{http://arxiv.org/abs/1512.08596}{{\tt arXiv:1512.08596}}].

\bibitem{Dai:2016gtz}
L.~Dai and K.~Fang, {\it {Can tidal disruption events produce the IceCube
  neutrinos?}},  \href{http://arxiv.org/abs/1612.00011}{{\tt
  arXiv:1612.00011}}.

\bibitem{Senno:2016bso}
N.~Senno, K.~Murase, and P.~M\'esz\'aros, {\it {High-energy neutrino flashes
  from x-ray bright and dark tidal disruptions events}},
  \href{http://arxiv.org/abs/1612.00918}{{\tt arXiv:1612.00918}}.

\bibitem{Kowalski2016}
M.~Kowalski, {\it {Talk presented at Neutrino 2016, XXVII International
  Conference on Neutrino Physics and Astrophysics, 4--9 July 2016, London
  (UK)}},  2016.

\bibitem{Aartsen:2016xlq}
{\bf IceCube} Collaboration, M.~G. Aartsen et~al., {\it {Observation and
  Characterization of a Cosmic Muon Neutrino Flux from the Northern Hemisphere
  using six years of IceCube data}},
  \href{http://arxiv.org/abs/1607.08006}{{\tt arXiv:1607.08006}}.

\bibitem{Aartsen:2016oji}
{\bf IceCube} Collaboration, M.~G. Aartsen et~al., {\it {All-sky search for
  time-integrated neutrino emission from astrophysical sources with 7 years of
  IceCube data}},  \href{http://arxiv.org/abs/1609.04981}{{\tt
  arXiv:1609.04981}}.

\bibitem{Aartsen:2016tpb}
{\bf IceCube} Collaboration, M.~G. Aartsen et~al., {\it {Lowering IceCube's
  Energy Threshold for Point Source Searches in the Southern Sky}},  {
  Astrophys. J.} {\bf 824} (2016), no.~2 L28,
  [\href{http://arxiv.org/abs/1605.00163}{{\tt arXiv:1605.00163}}].

\bibitem{Aartsen:2016qcr}
{\bf IceCube} Collaboration, M.~G. Aartsen et~al., {\it {An All-Sky Search for
  Three Flavors of Neutrinos from Gamma-Ray Bursts with the IceCube Neutrino
  Observatory}},  { Astrophys. J.} {\bf 824} (2016), no.~2 115,
  [\href{http://arxiv.org/abs/1601.06484}{{\tt arXiv:1601.06484}}].

\bibitem{Adrian-Martinez:2015ver}
{\bf IceCube, ANTARES} Collaboration, S.~Adrian-Martinez et~al., {\it {The
  First Combined Search for Neutrino Point-sources in the Southern Hemisphere
  With the Antares and Icecube Neutrino Telescopes}},  { Astrophys. J.} {\bf
  823} (2016), no.~1 65, [\href{http://arxiv.org/abs/1511.02149}{{\tt
  arXiv:1511.02149}}].

\bibitem{Aartsen:2015yva}
{\bf IceCube} Collaboration, M.~G. Aartsen et~al., {\it {The IceCube Neutrino
  Observatory - Contributions to ICRC 2015 Part I: Point Source Searches}},
  \href{http://arxiv.org/abs/1510.05222}{{\tt arXiv:1510.05222}}.

\bibitem{Ahlers:2014ioa}
M.~Ahlers and F.~Halzen, {\it {Pinpointing Extragalactic Neutrino Sources in
  Light of Recent IceCube Observations}},  { Phys. Rev.} {\bf D90} (2014),
  no.~4 043005, [\href{http://arxiv.org/abs/1406.2160}{{\tt arXiv:1406.2160}}].

\bibitem{Murase:2016gly}
K.~Murase and E.~Waxman, {\it {Constraining High-Energy Cosmic Neutrino
  Sources: Implications and Prospects}},  { Phys. Rev.} {\bf D94} (2016),
  no.~10 103006, [\href{http://arxiv.org/abs/1607.01601}{{\tt
  arXiv:1607.01601}}].

\bibitem{Feyereisen:2016fzb}
M.~R. Feyereisen, I.~Tamborra, and S.~Ando, {\it {One-point fluctuation
  analysis of the high-energy neutrino sky}},
  \href{http://arxiv.org/abs/1610.01607}{{\tt arXiv:1610.01607}}.

\bibitem{Halzen:2016seh}
F.~Halzen, A.~Kheirandish, and V.~Niro, {\it {Prospects for Detecting Galactic
  Sources of Cosmic Neutrinos with IceCube: An Update}},  { Astropart. Phys.}
  {\bf 86} (2017) 46--56, [\href{http://arxiv.org/abs/1609.03072}{{\tt
  arXiv:1609.03072}}].

\bibitem{Aartsen:2016lir}
{\bf IceCube} Collaboration, M.~G. Aartsen et~al., {\it {The contribution of
  Fermi-2LAC blazars to the diffuse TeV-PeV neutrino flux}},
  \href{http://arxiv.org/abs/1611.03874}{{\tt arXiv:1611.03874}}.

\bibitem{Aartsen:2014cva}
{\bf IceCube} Collaboration, M.~G. Aartsen et~al., {\it {Searches for Extended
  and Point-like Neutrino Sources with Four Years of IceCube Data}},  {
  Astrophys. J.} {\bf 796} (2014), no.~2 109,
  [\href{http://arxiv.org/abs/1406.6757}{{\tt arXiv:1406.6757}}].

\bibitem{Crook:2006sw}
A.~C. Crook, J.~P. Huchra, N.~Martimbeau, K.~L. Masters, T.~Jarrett, and L.~M.
  Macri, {\it {Groups of Galaxies in the Two Micron All-Sky Redshift Survey}},
  { Astrophys. J.} {\bf 655} (2007) 790--813,
  [\href{http://arxiv.org/abs/astro-ph/0610732}{{\tt astro-ph/0610732}}].

\bibitem{Aartsen:2014njl}
{\bf IceCube} Collaboration, M.~G. Aartsen et~al., {\it {IceCube-Gen2: A Vision
  for the Future of Neutrino Astronomy in Antarctica}},
  \href{http://arxiv.org/abs/1412.5106}{{\tt arXiv:1412.5106}}.

\bibitem{Murase:2012df}
K.~Murase, J.~F. Beacom, and H.~Takami, {\it {Gamma-Ray and Neutrino
  Backgrounds as Probes of the High-Energy Universe: Hints of Cascades, General
  Constraints, and Implications for TeV Searches}},  { JCAP} {\bf 1208} (2012)
  030, [\href{http://arxiv.org/abs/1205.5755}{{\tt arXiv:1205.5755}}].

\bibitem{Abbasi:2010rd}
{\bf IceCube} Collaboration, R.~Abbasi et~al., {\it {Time-Integrated Searches
  for Point-like Sources of Neutrinos with the 40-String IceCube Detector}},  {
  Astrophys. J.} {\bf 732} (2011) 18,
  [\href{http://arxiv.org/abs/1012.2137}{{\tt arXiv:1012.2137}}].

\bibitem{Fang:2016hyv}
K.~Fang and M.~C. Miller, {\it {A New Method for Finding Point Sources in
  High-energy Neutrino Data}},  { Astrophys. J.} {\bf 826} (2016), no.~2 102,
  [\href{http://arxiv.org/abs/1603.09306}{{\tt arXiv:1603.09306}}].

\bibitem{Yuksel:2008cu}
H.~Yuksel, M.~D. Kistler, J.~F. Beacom, and A.~M. Hopkins, {\it {Revealing the
  High-Redshift Star Formation Rate with Gamma-Ray Bursts}},  { Astrophys. J.}
  {\bf 683} (2008) L5--L8, [\href{http://arxiv.org/abs/0804.4008}{{\tt
  arXiv:0804.4008}}].

\bibitem{Gorski:2004by}
K.~M. Gorski, E.~Hivon, A.~J. Banday, B.~D. Wandelt, F.~K. Hansen, M.~Reinecke,
  and M.~Bartelman, {\it {HEALPix - A Framework for high resolution
  discretization, and fast analysis of data distributed on the sphere}},  {
  Astrophys. J.} {\bf 622} (2005) 759--771,
  [\href{http://arxiv.org/abs/astro-ph/0409513}{{\tt astro-ph/0409513}}].

\bibitem{Wilks:1938dza}
S.~S. Wilks, {\it {The Large-Sample Distribution of the Likelihood Ratio for
  Testing Composite Hypotheses}},  { Annals Math. Statist.} {\bf 9} (1938),
  no.~1 60--62.

\bibitem{Gruppioni:2013jna}
C.~Gruppioni et~al., {\it {The Herschel PEP/HerMES Luminosity Function. I:
  Probing the Evolution of PACS selected Galaxies to z~4}},  { Mon. Not. Roy.
  Astron. Soc.} {\bf 432} (2013) 23,
  [\href{http://arxiv.org/abs/1302.5209}{{\tt arXiv:1302.5209}}].

\bibitem{Ackermann:2012vca}
{\bf Fermi-LAT} Collaboration, M.~Ackermann et~al., {\it {GeV Observations of
  Star-forming Galaxies with \textit{Fermi} LAT}},  { Astrophys. J.} {\bf 755}
  (2012) 164, [\href{http://arxiv.org/abs/1206.1346}{{\tt arXiv:1206.1346}}].

\bibitem{Peng:2016nsx}
F.-K. Peng, X.-Y. Wang, R.-Y. Liu, Q.-W. Tang, and J.-F. Wang, {\it {First
  detection of GeV emission from an ultraluminous infrared galaxy: Arp 220 as
  seen with the Fermi Large Area Telescope}},  { Astrophys. J.} {\bf 821}
  (2016), no.~2 L20, [\href{http://arxiv.org/abs/1603.06355}{{\tt
  arXiv:1603.06355}}].

\bibitem{Griffin:2016wzb}
R.~D. Griffin, X.~Dai, and T.~A. Thompson, {\it {Constraining Gamma-Ray
  Emission from Luminous Infrared Galaxies with Fermi-LAT; Tentative Detection
  of Arp 220}},  { Astrophys. J.} {\bf 823} (2016), no.~1 L17,
  [\href{http://arxiv.org/abs/1603.06949}{{\tt arXiv:1603.06949}}].

\bibitem{Ajello:2011zi}
M.~Ajello et~al., {\it {The Luminosity Function of Fermi-detected Flat-Spectrum
  Radio Quasars}},  { Astrophys. J.} {\bf 751} (2012) 108,
  [\href{http://arxiv.org/abs/1110.3787}{{\tt arXiv:1110.3787}}].

\bibitem{Ajello:2013lka}
M.~Ajello et~al., {\it {The Cosmic Evolution of Fermi BL Lacertae Objects}},  {
  Astrophys. J.} {\bf 780} (2014) 73,
  [\href{http://arxiv.org/abs/1310.0006}{{\tt arXiv:1310.0006}}].

\bibitem{Inoue:2011bm}
Y.~Inoue, {\it {Contribution of the Gamma-ray Loud Radio Galaxies Core
  Emissions to the Cosmic MeV and GeV Gamma-Ray Background Radiation}},  {
  Astrophys. J.} {\bf 733} (2011) 66,
  [\href{http://arxiv.org/abs/1103.3946}{{\tt arXiv:1103.3946}}].

\bibitem{Tjus:2014dna}
J.~Becker~Tjus, B.~Eichmann, F.~Halzen, A.~Kheirandish, and S.~M. Saba, {\it
  {High-energy neutrinos from radio galaxies}},  { Phys. Rev.} {\bf D89}
  (2014), no.~12 123005, [\href{http://arxiv.org/abs/1406.0506}{{\tt
  arXiv:1406.0506}}].

\bibitem{Becker:2005ya}
J.~K. Becker, P.~L. Biermann, and W.~Rhode, {\it {The Diffuse neutrino flux
  from FR-II radio galaxies and blazars: A Source property based estimate}},  {
  Astropart. Phys.} {\bf 23} (2005) 355--368,
  [\href{http://arxiv.org/abs/astro-ph/0502089}{{\tt astro-ph/0502089}}].

\bibitem{Kimura:2014jba}
S.~S. Kimura, K.~Murase, and K.~Toma, {\it {Neutrino and Cosmic-Ray Emission
  and Cumulative Background from Radiatively Inefficient Accretion Flows in
  Low-Luminosity Active Galactic Nuclei}},  { Astrophys. J.} {\bf 806} (2015)
  159, [\href{http://arxiv.org/abs/1411.3588}{{\tt arXiv:1411.3588}}].

\end{thebibliography}\endgroup

\end{document}